\documentclass[a4paper,UKenglish,cleveref, autoref, thm-restate, numberwithinsect]{lipics-v2021}

\pdfoutput=1 %
\hideLIPIcs  %

\graphicspath{{./graphics-cgta/}}%

\title{Constrained Boundary Labeling} %

\author{Thomas Depian}{Algorithms and Complexity Group, TU Wien, Austria}{tdepian@ac.tuwien.ac.at}{https://orcid.org/0009-0003-7498-6271}{Vienna Science and Technology Fund (WWTF) [10.47379/ICT22029].}%

\author{Martin N\"ollenburg}{Algorithms and Complexity Group, TU Wien, Austria}{noellenburg@ac.tuwien.ac.at}{https://orcid.org/0000-0003-0454-3937}{}

\author{Soeren Terziadis}{Algorithms cluster, TU Eindhoven, The Netherlands}{s.d.terziadis@tue.nl}{https://orcid.org/0000-0001-5161-3841}{Vienna Science and Technology Fund (WWTF) [10.47379/ICT19035] and European Union's Horizon 2020 research and innovation programme under the Marie Skłodowska-Curie grant agreement No 101034253.}

\author{Markus Wallinger}{Chair for Efficient Algorithms, Technical University of Munich, Germany}{markus.wallinger@tum.de}{https://orcid.org/0000-0002-2191-4413}{Vienna Science and Technology Fund (WWTF) [10.47379/ICT19035].}

\authorrunning{T.\, Depian, M.\, N\"ollenburg, S.\, Terziadis, M.\, Wallinger} %

\Copyright{Thomas Depian, Martin N\"ollenburg, Soeren Terziadis, Markus Wallinger} %



\category{} %

\relatedversiondetails[cite={CGTA},linktext={\\https://doi.org/10.1016/j.comgeo.2025.102191}]{Accepted for publication in Computational Geometry}{https://doi.org/10.1016/j.comgeo.2025.102191} %

\nolinenumbers %

\EventEditors{Juli\'{a}n Mestre and Anthony Wirth}
\EventNoEds{2}
\EventLongTitle{35th International Symposium on Algorithms and Computation (ISAAC 2024)}
\EventShortTitle{ISAAC 2024}
\EventAcronym{ISAAC}
\EventYear{2024}
\EventDate{December 8--11, 2024}
\EventLocation{Sydney, Australia}
\EventLogo{}
\SeriesVolume{322}
\ArticleNo{42}

\usepackage{xspace}
\usepackage[table]{xcolor}
\usepackage{todonotes}
\usepackage{pifont}
\usepackage{layout}
\usepackage{booktabs}
\usepackage{mathtools}
\usepackage{multirow}
\usepackage[hyphenbreaks]{breakurl}

\newcommand{\Sites}{\ensuremath{\mathcal{S}}\xspace}

\newcommand{\Candidates}{\ensuremath{\mathcal{C}}\xspace}
\newcommand{\Instance}{\ensuremath{\mathcal{I}}\xspace}
\newcommand{\Labeling}{\ensuremath{\mathcal{L}}\xspace}
\newcommand{\Label}{\ensuremath{\ell}\xspace}
\newcommand{\Leader}{\ensuremath{\lambda}\xspace}
\newcommand{\Leaders}{\ensuremath{\Lambda}\xspace}
\newcommand{\Boundary}{\ensuremath{\mathcal{B}}\xspace}
\newcommand{\Order}{\ensuremath{\preccurlyeq}}
\newcommand{\OrderingConstraints}{\ensuremath{\preccurlyeq}\xspace}
\newcommand{\Group}{\ensuremath{\mathcal{G}}\xspace}
\newcommand{\GroupingConstraints}{\ensuremath{\Gamma}\xspace}%

\newcommand{\ConstraintsLong}{\ensuremath{\left(\GroupingConstraints, \OrderingConstraints\right)}\xspace}
\newcommand{\PQAGraph}{\ensuremath{\mathcal{T}}\xspace}
\newcommand{\PQTree}{\ensuremath{\tau\xspace}}

\newcommand{\Restricted}[2]{{#1}(#2)}

\newcommand{\LeaderStyle}[1]{\ensuremath{\mathit{#1}}}
\newcommand{\Po}{\LeaderStyle{po}}

\newcommand{\Leaves}[1]{\ensuremath{\text{leaves}(#1)}}

\newcommand{\BigO}[1]{\ensuremath{\mathcal{O}(#1)}}

\newcommand{\NP}{\textsf{NP}}

\newcommand{\probname}[1]{{\normalfont\textsc{#1}}}

\newcommand{\KSCBLProblem}[1]{\probname{Constrained {#1}-Sided Boundary Labeling}\xspace}
\newcommand{\KSCBLProblemShort}[1]{\probname{{#1}-CBL}\xspace}

\newcommand{\PositiveOneThreeSat}{\probname{Positive 1-In-3 Sat}\xspace}

\newcommand{\KSCBLProblemHeader}[1]{Constrained {#1}-Sided Boundary Labeling\xspace}

\newcommand{\RespectsConstraints}[1]{\ensuremath{\probname{RespectsConstraints}(#1)}}
\newcommand{\Admissible}[1]{\ensuremath{\probname{Admissible}(#1)}}

\newcommand{\SolverILP}{Na\"ive ILP\xspace}

\newcommand{\Size}[1]{\ensuremath{\left\vert #1 \right\vert}}

\newtheorem*{problem*}{Problem}

\newif\ifcomments
\commentstrue
\definecolor{Group1}{RGB}{26, 157, 118}
\definecolor{Group2}{RGB}{216, 94, 1}

\newcommand{\admissibleText}{admissible}

\newcommand{\portText}{port}
\newcommand{\portsText}{ports}

\newcommand{\referencepointText}{reference point}
\newcommand{\referencepointsText}{reference points}

\newcommand{\ReferencePointsText}{Reference Points}

\newcommand{\candidateText}{candidate}
\newcommand{\candidatesText}{candidates}
\newcommand{\CandidateText}{Candidate}
\newcommand{\CandidatesText}{Candidates}

\newcommand{\rootOfTree}[1]{\ensuremath{\text{root}(#1)}}

\newcommand{\NewText}[1]{#1}

\begin{document}

\maketitle

\begin{abstract}
Boundary labeling is a technique in computational geometry used to label %
sets of features in an illustration. 
It involves placing labels along an axis-parallel %
bounding box and connecting each label with its corresponding feature %
using non-crossing leader lines.
Although boundary labeling is well-studied, semantic constraints on the labels have not been investigated thoroughly.
In this paper, we introduce \emph{grouping} and \emph{ordering constraints} in boundary labeling:
Grouping constraints enforce that all labels in a group are placed consecutively on the boundary, and ordering constraints enforce a partial order over the labels.
We show that it is %
\NP-hard to find a labeling for arbitrarily sized labels with unrestricted positions along one side of the boundary.
However, we obtain polynomial-time algorithms if we restrict this problem either to uniform-height labels or to a finite set of candidate positions. 
Furthermore, we show that finding a labeling on two opposite sides of the boundary is \NP-complete, even for uniform-height labels and finite label positions.
Finally, we experimentally confirm that our approach has also practical relevance.
\end{abstract}

\section{Introduction}
\begin{figure}[t]
	\centering
	\begin{subfigure}[t]{.45\linewidth}
		\centering
		\includegraphics[page=1]{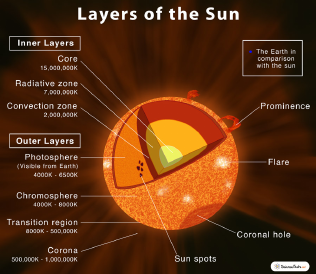}
		\subcaption{Schematic of the sun.~\copyright\space ScienceFacts.net~\cite{Bhuyan.2023}; reproduced with permission.}
		\label{fig:illustration-motivation_a}
	\end{subfigure}
	\hfill
	\begin{subfigure}[t]{.45\linewidth}
		\centering
		\includegraphics[page=2,scale=.9]{figure_01}
		\subcaption{Cities in Italy. Labeling with \Po-leaders created by our algorithm described in \cref{sec:fixed-ports}.}
		\label{fig:illustration-motivation_b}
	\end{subfigure}
	\caption{Labelings that adhere to semantic constraints.}
	\label{fig:illustration-motivation}
\end{figure}

Annotating features of interest with textual information in illustrations, e.g.,  in technical, medical, or geographic domains, is an important and challenging task in graphic design and information visualization.
One common guideline when creating such labeled illustrations is to ``not obscure important details with labels''~\cite[p.~35]{Briscoe.1990}.
Therefore, for complex illustrations, designers tend to place the labels outside the illustrations, creating an \emph{external labeling} as shown in \cref{fig:illustration-motivation_a}.
Feature points, called \emph{sites}, are connected to descriptive labels with non-crossing polyline \emph{leaders}, while optimizing an objective function, e.g., the total leader length.

External labeling is a well-studied area both from a practical visualization perspective~\cite{Battersby2011} and from a formal algorithmic perspective~\cite{Bekos.2021}.
One aspect of external labeling that has not yet been thoroughly studied in the literature is that of constraining the placement of (subsets of) labels in the %
creation process.
The sequential arrangement of external labels along the boundary of the illustration creates %
spatial proximities between the labels that do not necessarily correspond to the geometric proximity patterns of the sites in the illustration.
Hence, it is of interest in many applications to put constraints on the grouping and ordering of these labels in order to improve the readability and semantic coherence of the labeled illustration.
Examples of such constraints could be to group labels of semantically related sites or to restrict the top-to-bottom order of certain labels to reflect some ordering of their sites in the illustration; see \cref{fig:illustration-motivation_a}, where the inner and outer layers of the sun are grouped and ordered from the core to the surface.

More precisely, we study such constrained labelings in the  \emph{boundary labeling} model, which is a well-studied special case of external labelings. Here, the labels must be placed along a rectangular boundary around the illustration~\cite{Bekos.2007}.
Initial work placed the labels on one or two sides of the boundary, usually the left and right sides.
For uniform-height labels, polynomial-time algorithms to compute a labeling that minimizes the length of the leaders~\cite{Bekos.2007} or more general optimization functions~\cite{Benkert.2009} have been proposed.
Polynomial-time approaches to compute a labeling with equal-sized labels on (up to) all four sides of the boundary are also known~\cite{Kindermann.2016}. 
For non-uniform height labels, %
\NP-hardness has been shown in the general two-sided~\cite{Bekos.2007}, and in different one-sided settings~\cite{Bekos.2010a,Fink.2016}.
Several leader styles have been considered, and we refer to the book of Bekos, Niedermann, and Nöllenburg~\cite{Bekos.2021} and the user study of Barth, Gemsa, Niedermann, and Nöllenburg~\cite{Barth.2019} for an overview.
We will focus on a frequently used class of L-shaped leaders, called \emph{\Po-leaders}, see also \cref{fig:illustration-motivation_b}, that consist of two segments: one is  \textbf{\emph{p}}arallel and the other \textbf{\emph{o}}rthogonal to the side of the boundary on which the label is placed~\cite{Bekos.2007}. %
These \Po-leaders turned out as the recommended leader type in the study of Barth et al.~\cite{Barth.2019} as they performed well in various readability tasks and received high user preference ratings.

The literature considered various extensions of boundary labeling~\cite{Bekos.2015,Fink.2016,Gedicke.2021,Gemsa.2015}, and we broaden this body of work with our paper that aims at systematically investigating the above-mentioned constraints in boundary labeling from an algorithmic perspective.

\subparagraph{Problem Description.}
In the following, we use the taxonomy of Bekos et al.~\cite{Bekos.2021} where applicable. 
Let \Sites be a set of $n$ sites in $\mathbb{R}^2$ enclosed in an axis-parallel bounding box $\Boundary$ and in general position, i.e., no two sites share the same~$x$- or $y$-coordinate.
For each site $s_i \in \Sites$, we have an open rectangle $\Label_i$ of height $h(\Label_i) > 0$ and some width, which we call the \emph{label} of the site.
The rectangles model the (textual) labels, which are usually a single line of text in a fixed font size, as their bounding boxes.
Hence, we often restrict ourselves to uniform-height labels, but neglect their width.
The \Po-leader $\Leader_i = (s_i, c_i)$ is a polyline consisting of (up to) one vertical and one horizontal segment and connects the site~$s_i$ with the \emph{reference point} $c_i$, which is a point on the boundary~\Boundary at which we attach the label $\Label_i$.
The point where $\Leader_i$ touches~$\Label_i$ is called the \emph{\portText{}} of $\Label_i$.
We define the \portText{} for each label \Label to be at half its height, i.e., we use \emph{fixed} ports.
Hence, the position of $c_i$ uniquely determines the position of the \portText{}~$p_i$ and thus the placement of the label $\Label_i$.
We let $\Leaders$ denote the set of all possible leaders and%
~$\Candidates$ the set of all possible reference points.
In a $b$-sided boundary labeling $\Labeling\colon \Sites \to \Candidates$, we route for each site $s \in \Sites$ a leader \Leader to the \referencepointText{} $\Labeling(s)$ on the right ($b = 1$) or the right and left ($b = 2$) side of \Boundary, such that we can (in a post-processing step) place the label~\Label for $s$ at the \referencepointText{} $\Labeling(s)$
and no two labels %
overlap. 
\begin{figure}[t]
	\centering
	\begin{subfigure}[t]{.21\linewidth}
		\centering
		\includegraphics[page=1]{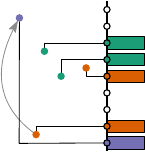}
		\subcaption{Length-minimal}
		\label{fig:labeling-examples_a}
	\end{subfigure}
	\hfill
	\begin{subfigure}[t]{.21\linewidth}
		\centering
		\includegraphics[page=2]{figure_02}
		\subcaption{Bend-minimal}
		\label{fig:labeling-examples_b}
	\end{subfigure}
	\hfill
	\begin{subfigure}[t]{.26\linewidth}
		\centering
		\includegraphics[page=4]{figure_02}
		\subcaption{A $2$-sided labeling}
		\label{fig:labeling-examples_c}
	\end{subfigure}
	\hfill
	\begin{subfigure}[t]{.21\linewidth}
		\centering
		\includegraphics[page=3]{figure_02}
		\subcaption{Non-\admissibleText}
		\label{fig:labeling-examples_d}
	\end{subfigure}
	\caption{
		Colors indicate grouping and arrows ordering constraints.
		Labelings that are optimal with respect to~(\subref{fig:labeling-examples_a}) the total leader length and~(\subref{fig:labeling-examples_b}) the number of bends. 
		In the $2$-sided layout~(\subref{fig:labeling-examples_c}) the ordering constraint $s_i \Order s_j$ is not enforced since $\ell_i$ and $\ell_j$ are on different sides.
		Note that~(\subref{fig:labeling-examples_d}) is a planar but non-\admissibleText{} length-minimal labeling as it violates both the ordering and the orange and green grouping constraints.
	} 
	\label{fig:labeling-examples}
\end{figure}
If \Candidates consists of a finite set of~$m$ candidate \referencepointsText{} on~\Boundary, we say that we have \emph{fixed (candidate) \referencepointsText{}}, otherwise \Candidates consists of the respective side(s) of \Boundary, which is called \emph{sliding (candidate) \referencepointsText{}}.
In the following, we call the \referencepointsText{} in \Candidates simply \emph{\candidatesText}.
A labeling is called \emph{planar} if no two labels overlap and there is no leader-leader or leader-site crossing.
We can access the $x$- and $y$-coordinate of a site, \portText{}, or \candidateText{} with $x(\cdot)$ and~$y(\cdot)$.

In our constrained boundary labeling setting, we are in addition given a tuple of constraints $\ConstraintsLong$ consisting of a family of $k$ grouping constraints $\GroupingConstraints = \{\Group_1, \ldots, \Group_k\}$ 
and a partial order \OrderingConstraints on the sites.
A \emph{grouping} constraint $\emptyset \ne \Group \subseteq \Sites$ enforces that the labels for the sites in $\Group$ appear consecutively on the same side of the boundary, as in \cref{fig:labeling-examples_a,fig:labeling-examples_b,fig:labeling-examples_c}, but in general there can be gaps between two labels of the same group; compare \cref{fig:labeling-examples_a,fig:labeling-examples_b}.
An \emph{ordering} constraint $s_i \Order s_j$ enforces for the labeling \Labeling to have $y(\Labeling(s_i)) \geq y(\Labeling(s_j))$\footnote{We assume that the lower-left corner of the boundary coincides with the origin of the coordinate system.} but only if~$s_i$ and $s_j$ are labeled on the same side of \Boundary.
Hence, if the labels $\Label_i$ and $\Label_j$ are placed on the same side of \Boundary, then the ordering constraint enforces that %
$\Label_j$ must not appear above %
$\Label_i$. %
On the other hand, if $\Label_i$ and $\Label_j$ are on different sides of \Boundary, then the constraint $s_i \Order s_j$ has no effect; %
see also \cref{fig:labeling-examples_c}. %
This interpretation is motivated by the Gestalt principle of proximity~\cite{Ware2013}, as the spatial distance between labels on opposite sides is usually large and we thus perceive labels on the same side as belonging together.
Furthermore, this interpretation of ordering constraints has already been applied in hand-made labelings~\cite{OrderingConstraints.2021}.
Note that in general \OrderingConstraints may contain reflexive %
constraints, which %
are fulfilled by any labeling and can thus be removed. The one-sided model in addition implicitly fulfills any transitive constraint.
Hence, we work in the one-sided model with the transitive reduction of $(\Sites, \OrderingConstraints)$. %
In the following, we let~$r$ denote the number of %
ordering constraints in the respective model.

A labeling \emph{respects} the grouping (ordering) constraints if all the grouping (ordering) constraints are satisfied.
Furthermore, the grouping (ordering) constraints are \emph{consistent} if there exists a (not necessarily planar) labeling that respects them.
Similarly, the constraints $\ConstraintsLong$ are consistent if there exists a labeling that respects \GroupingConstraints and \OrderingConstraints simultaneously.
Finally, a labeling is \emph{admissible} if it is planar and respects the constraints.
Observe that an instance with inconsistent constraints can never have an \admissibleText{} labeling.
If an \admissibleText{} labeling exists, we want to optimize a quality criterion expressed by a function $f\colon \Leaders \to \mathbb{R}_0^+$.
In this paper, $f$ measures the length or the number of bends of a leader, which are the most commonly used criteria~\cite{Bekos.2021}. 
\cref{fig:labeling-examples} highlights the differences and shows with \cref{fig:labeling-examples_d} that an optimal \admissibleText{} labeling might be, with respect to $f$, worse than its planar (non-admissible) counterpart. 
We are now ready to define our problem of interest.
\begin{problem*}
	\KSCBLProblem{$b$} (\KSCBLProblemShort{$b$}):
	\begin{description}
		\item[Given:] An instance \Instance consisting of a set of $n$ sites $\Sites = \{s_1, \ldots, s_n\}$, a set of~$n$ labels $L = \{\Label_1, \ldots, \Label_n\}$, a function $h\colon L \to \mathbb{R}^+$ that assigns to the label~$\Label_i$ for the site~$s_i$ a positive height $h(\Label_i) \in \mathbb{R}^+$, a quality function $f\colon \Leaders \to \mathbb{R}_0^+$, an axis-parallel bounding box~$\Boundary$, a set of $k$ grouping constraints $\GroupingConstraints = \{\Group_1, \ldots, \Group_k\}$, a partial order~\OrderingConstraints on the sites consisting of~$r$ constraints in the transitive reduction of $(\Sites, \OrderingConstraints)$, and optionally a set of~$m$ \candidatesText{} \Candidates on $b$ sides of \Boundary.
		\item[Task:] Find an \admissibleText{} $b$-sided \Po-labeling $\Labeling^*$ for \Instance (on the \candidatesText{} \Candidates) that minimizes $\sum_{s \in \Sites} f((s, \Labeling^*(s))$ or report that no \admissibleText{} labeling exists.
	\end{description}
\end{problem*}

\subparagraph{Related Work on Constrained External Labeling.}
Our work is in line with (recent) efforts to integrate semantic constraints into the external labeling model.
The survey of Bekos et al.~\cite{Bekos.2021} reports papers that group labels.
Some results consider heuristic label placements in interactive 3D  visualizations~\cite{Goetzelmann.2006,Goetzelmann.2006a} or group (spatially close) sites together to label them with a single label~\cite{Fink.2012,Muehler2009,Tatzgern13}, bundle the leaders~\cite{Niedermann2019}, or align the labels~\cite{Vollick.2007}.
These papers are usually targeted at applications and do not aim for exact algorithms or formal complexity bounds.
Moreover, the grouping is often not part of the input but rather determined by clustering similar sites or sites that are located close to each other.
To the best of our knowledge, Niedermann, Nöllenburg, and Rutter~\cite{Niedermann.2017} are the first that support the explicit grouping of labels while ensuring a planar labeling.
They proposed a contour labeling algorithm, a generalization of boundary labeling, that can be extended to group sets of labels as hard constraints in their model.
However, they did not analyze this extension in detail and do not support ordering constraints. 
Recently, Gedicke, Arzoumanidis, and Haunert~\cite{Gedicke.2023} tried to maximize the number of respected groups that arise from the spatial proximity of the sites or their semantics, i.e., they reward a labeling also based on the number of consecutive labels from the same group. 
They disallow assigning a site to more than one group, but see combining spatial and semantic groups as an interesting direction for further research.
We work towards that goal, as we allow grouping constraints to overlap.
Finally, Klawitter, Klesen, Scholl, and van Dijk~\cite{Klawitter.2023} visualized geophylogenies by embedding a binary (phylogenetic) tree on one side of the boundary.
Each leaf of the tree corresponds to a site, and the aim is to connect them using straight-line leaders with few crossings.
These trees implicitly encode grouping constraints, as sites with a short path between their leaves must be labeled close together on the boundary.
However, Klawitter et al.\ not only considered a different optimization function, but restricted themselves to binary trees, which cannot represent all grouping constraints.
Overall, we can identify a growing interest in grouping and ordering constraints in the literature, but a systematic investigation of these constraints from an algorithmic perspective is still lacking.
This paper should be a step towards filling this gap by allowing arbitrary grouping and ordering constraints as input.

\subparagraph{Contributions.}
In \cref{sec:one-sided}, we take a closer look at \KSCBLProblemShort{1}.
We prove that it is %
\NP-hard to find an \admissibleText{} labeling with sliding \candidatesText{} and unrestricted label heights (\cref{sec:one-sided-hardness})
and present polynomial-time algorithms for fixed \candidatesText{} and unrestricted label heights (\cref{sec:fixed-ports}) and for sliding \candidatesText{} and uniform-height labels (\cref{sec:sliding-ports}).
To that end, we combine a dynamic program with a novel data structure based on PQ-trees.
We show in \cref{sec:two-sided} that \KSCBLProblemShort{2} is \NP-complete, even for uniform-height labels and fixed \candidatesText{}.
To the best of our knowledge, this is the first two-sided boundary labeling problem that is already \NP-hard in such a restricted setting.
We summarize our theoretical results in \cref{tab:results}.
\cref{tab:results} makes immanent that our algorithms, although being polynomial, have a high running time. 
To complement our theoretic results, we implemented our dynamic program for fixed \candidatesText{} and report in \cref{sec:experiments} an experimental evaluation of its performance for realistically sized real-world instances.
These experiments underline that our algorithm for fixed \candidatesText{} has also practical relevance.
\begin{table}
	\small
	\centering
	\caption{Our results on \KSCBLProblemShort{$b$} with $n$ sites, $m$ candidates, and a family \GroupingConstraints of $k$ grouping constraints.}
	\begin{tabular}{cccrr} 
		\toprule
		$b$ & candidates & label height & result & reference \\ 
		\midrule 
		1 & sliding & non-uniform & %
		\NP-hard & \cref{thm:hardness-1-sided}\\
		1 & fixed & non-uniform & \BigO{n^4m^3\log m + k + \sum_{\Group \in \GroupingConstraints} \Size{\Group}} & \cref{thm:dp-running-time-space}\\
		1 & sliding & uniform & $\BigO{n^{10}\log n + k + \sum_{\Group \in \GroupingConstraints} \Size{\Group}}$ & \cref{thm:sliding-running-time-space}\\
		\midrule
		2 & fixed & uniform & \NP-complete & \cref{thm:hardness-2-sided}\\
		\bottomrule
	\end{tabular}
	\label{tab:results}
\end{table}

\section{The \KSCBLProblemHeader{One} Problem}
\label{sec:one-sided}
We start with investigating the computational complexity of \KSCBLProblemShort{1}.
In \cref{sec:one-sided-hardness}, we show that finding an \admissibleText{} labeling for an instance of \KSCBLProblemShort{1} is %
\NP-hard.%
However, by restricting the input to either fixed \candidatesText{} (\cref{sec:fixed-ports}) or uniform-height labels (\cref{sec:sliding-ports}), we can obtain polynomial-time algorithms.

\subsection{\KSCBLProblemHeader{One} is %
	\NP-hard}
\label{sec:one-sided-hardness}
\looseness=-1
To show that it is %
\NP-hard to find an \admissibleText{} labeling in an instance of \KSCBLProblemShort{1}, we can reduce from \probname{Partition}.
Inspired by a reduction by Fink and Suri~\cite{Fink.2016} from \probname{Partition} to the problem of finding a planar labeling with non-uniform height labels and sliding \candidatesText{} in the presence of a single obstacle on the plane, we %
create a gadget of sites with grouping and ordering constraints that simulates such an obstacle.

\subparagraph*{Construction of the Instance.}
We %
first give a reduction that only uses grouping constraints.
In the end, we show how we can replace the grouping constraints by ordering constraints.
Let $\left(\mathcal{A} = \{a_1, \ldots a_N\}, \omega\colon \mathcal{A} \to \mathbb{N}^+\right)$ be an instance of the weakly \NP-complete problem \probname{Partition} that is defined as follows~\cite{Garey.1979}.
Given a set $\mathcal{A}$ of $N$ elements together with a weight-function $\omega$ that assigns each element $a\in\mathcal{A}$ a non-negative weight $w(a)$; does there exist a subset $\mathcal{A}' \subseteq \mathcal{A}$ such that $\sum_{a \in \mathcal{A}'} \omega(a) = \sum_{a \in \mathcal{A} \setminus \mathcal{A}'} \omega(a) =~A$, for some integer~$A$?
We can assume $A \geq 6$.
This is without loss of generality, as we could otherwise solve the instance in polynomial time and return a trivial positive or negative instance of \KSCBLProblemShort{1}.

\begin{figure}
	\centering
	\includegraphics[page=1]{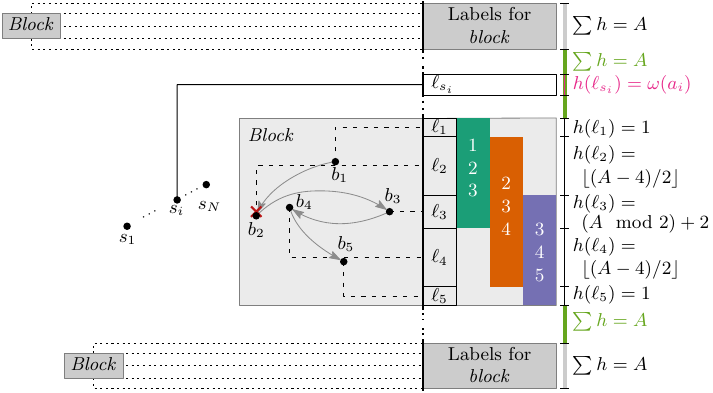}
	\caption{Components of the instance created by our %
		\NP-hardness reduction.
		Colored bands visualize the grouping constraints of a block, that can be replaced by ordering constraints (see the arrows). Note that the label heights and distances in this figure are not to scale. We indicate the anchor of the central block with the red cross.}
	\label{fig:non-uniform-hardness-block}
\end{figure}

In our reduction, we create one site $s_i$ for each element $a_i \in \mathcal{A}$ with a label of height $\omega(a_i)$.
We introduce five additional sites that admit, thanks to their label height paired with grouping constraints, only a single labeling at a predetermined position on the boundary.
These labels divide the boundary into two parts and by placing the sites for the elements of $\mathcal{A}$ behind these labels, we can mimic the partition process of $\mathcal{A}$.
To ensure that each part can only occupy labels whose heights sum up to $A$, we introduce two copies of the above structure that encapsulate the whole construction and leave only two $A$-high free windows on the boundary, corresponding to a partition of $\mathcal{A}$ into $\mathcal{A}'$ and $\mathcal{A} \setminus \mathcal{A}'$.
In the following, we give a detailed construction of the above high-level description which we visualize in \cref{fig:non-uniform-hardness-block}.

Without loss of generality, we can assume $N \geq 2$, as $\left(\mathcal{A}, \omega\right)$ would otherwise be a trivial negative instance.
Hence, $2\varepsilon < 1$ holds.
We start with describing the construction of the mentioned obstacle, which %
we call a \emph{block}, and create five sites,~$b_1$ to~$b_5$, with corresponding labels of height 1 for $b_1$ and $b_5$, height $\lfloor (A - 4) / 2\rfloor$ for~$b_2$ and~$b_4$, and height 2 or 3 for $b_3$, depending on whether $A$ is even or odd, respectively.
Observe that the heights of the labels for $b_1$ to $b_5$ sum up to~$A$ %
as also indicated in \cref{fig:non-uniform-hardness-block}.
In the course of the reduction, we create three blocks.
Thus, we now give a general placement of the sites of a block $B$ with respect to the \emph{anchor} point $(x_B, y_B)$ of the block.
Furthermore, to ensure general position of the construction while at the same time limiting the movement of the labels (and thus their position) to a distance of at most one, we introduce $\varepsilon = 1 / (2N)$.
For an anchor $(x_B, y_B)$, we place the sites $b_1$ to $b_5$ at $b_1 = (x_B + 2 - \varepsilon, y_B + h(\Label_3) / 2 + h(\Label_2) / 2 + \varepsilon)$, $b_2 = (x_B, y_B - \varepsilon)$, $b_3 = (x_B + 3, y_B)$, $b_4 = (x_B + 1, y_B + \varepsilon)$, and $b_5 = (x_B + 2 +~\varepsilon, y_B - h(\Label_3) / 2 - h(\Label_4) / 2 -\varepsilon)$.
Observe that the sites $b_1$ to $b_5$ are in general position.
A block contains the grouping constraints $\{b_1, b_2, b_3\}$, $\{b_2, b_3, b_4\}$, and $\{b_3, b_4, b_5\}$, which enforce that any \admissibleText{} labeling must label these sites as indicated in \cref{fig:non-uniform-hardness-block}.
Furthermore, the labels $\Label_2$, $\Label_3$, and $\Label_4$ can slide along the boundary at most by a total of $2 \varepsilon$.
However, recall that $2\varepsilon < 1$ holds.
As any label that we %
introduce has a height of at least one, the vertical distance these labels can slide is smaller than the height of any label.
Hence, we can neglect this sliding in the upcoming arguments and consider their placement as fixed, i.e., the labels $\Label_2$, $\Label_3$, and $\Label_4$ occupy an (almost) contiguous block on the boundary at a fixed position.
We now place a block on the sites~$b_1$ to~$b_5$ with anchor $(9 + N, 5A / 2)$ and two additional blocks, one on the sites~$x_1$ to $x_5$ with anchor $(1, 5A - A/2)$, and one on the sites $y_1$ to $y_5$ with anchor $(5, A/2)$.
Observe in \cref{fig:non-uniform-hardness-block} that these three blocks induce two $A$-high free windows on the boundary. %
These windows %
are the only place the labels for the sites that we introduced next can be placed.
More concretely, for each $a_i \in \mathcal{A}$, $1 \leq i \leq \lfloor N/2 \rfloor$, we place a site $s_i = (8 + i, 5A / 2 - \varepsilon(\lfloor N/2\rfloor - i + 2))$ and for each $a_i \in \mathcal{A}$, $\lfloor N/2 \rfloor < i \leq N$, we place a site $s_i = (8 + i, 5A / 2 + \varepsilon(i - \lfloor N/2\rfloor + 1))$.
Note that these sites are either below the site $b_2$ or above the site $b_4$ of the middle block.
Furthermore, these sites are left of the above-mentioned block and right of the other two blocks.
Thus, we ensure that our created instance %
is in general position.
The label~$\Label_{s_i}$ for the site $s_i$ has a height of $h(\Label_{s_i}) = \omega(a_i)$.
Recall that our construction left two $A$-high free windows on the boundary between the blocks and observe that the sites for the elements of $\mathcal{A}$ are placed entirely behind the (enforced) label positions for the middle block.
Hence, these windows are the only place where the labels for the sites representing the elements of~$\mathcal{A}$ can be.
Thus, when creating an \admissibleText{} labeling, one has to decide for each $s_i$ whether to label it above or below the block.
This can be used to form an equivalence between partitioning the elements of $\mathcal{A}$ into two sets, $\mathcal{A}_1$ and $\mathcal{A}_2$, and placing the labels for the sites in the upper or lower window on the boundary, respectively.

Finally, note that the grouping constraints we used in the blocks can be exchanged by the ordering constraints $\{b_1 \Order b_2, b_2 \Order b_3, b_3 \Order b_4, b_4 \Order b_5\}$ (see the arrows in \cref{fig:non-uniform-hardness-block}).
Similar substitutions can also be performed in the other two blocks.
We are now ready to show \cref{thm:hardness-1-sided}.

\begin{restatable}{theorem}{hardnessOneSidedTheorem}
	\label{thm:hardness-1-sided}
	Deciding if an instance of \KSCBLProblemShort{1} has an \admissibleText{} labeling is %
	\NP-hard, even for a constant number of grouping or ordering constraints.
\end{restatable}
\begin{proof}
	Let $\left(\mathcal{A} = \{a_1, \ldots a_N\}, \omega\right)$ be an instance of \probname{Partition} and let $\Instance\left(\mathcal{A}, \omega\right)$ be the instance of \KSCBLProblemShort{1} obtained with the construction from above.
	A close analysis of the construction reveals that $\Instance(\mathcal{A}, \omega)$ has $N + 15$ sites and $18$ grouping constraints (or 12 ordering constraints).
	The area occupied by $\Instance\left(\mathcal{A}, \omega\right)$ is in \BigO{AN}.
	Furthermore, scaling the instance by $2N$ would lead to $\varepsilon = 1$, thus allowing us to embed the instance on an integer grid of polynomial size.
	Clearly, $\Instance\left(\mathcal{A}, \omega\right)$ can be constructed in polynomial time.
	Thus, it remains to show the correctness of the reduction.
	
	\proofsubparagraph*{($\boldsymbol{\Rightarrow}$)}
	Let $\left(\mathcal{A}, w\right)$ be a positive instance of \probname{Partition} and $\mathcal{A}_1$ and $\mathcal{A}_2$ a corresponding solution, i.e., we have $\sum_{a \in \mathcal{A}_1} w(a) = \sum_{a \in \mathcal{A}_2} w(a) = A$.
	We transform this solution now into an \admissibleText{} labeling for $\Instance(\mathcal{A}, \omega)$.
	The blocks are labeled as in \cref{fig:non-uniform-hardness-block}.
	To determine the position of the labels for the remaining sites, $s_1$ to $s_N$, we traverse them from left to right.
	For any site $s_i$, $1 \leq i \leq N$, we check whether its corresponding element $a_i$ is in $\mathcal{A}_1$ or $\mathcal{A}_2$.
	If $a_i \in \mathcal{A}_1$, we place the label for $s_i$ as far up as possible, i.e., inside the upper window; recall \cref{fig:non-uniform-hardness-block}.
	On the other hand, if $a_i \in \mathcal{A}_2$, we place the label as far down as possible, i.e., inside the lower window.
	As the sum of the weights of the elements of $\mathcal{A}_1$ and $\mathcal{A}_2$ evaluates to $A$, respectively, and we reflect the values of the sum in the height of the respective labels, it is guaranteed that we can place all labels for sites that represent entries in $\mathcal{A}_1$ and $\mathcal{A}_2$ in the respective $A$-high windows on the boundary.
	Finally, by traversing the sites from left to right and assigning the outermost possible position for the respective label, we guarantee that the resulting labeling is planar and thus \admissibleText{}.
	
	\proofsubparagraph*{($\boldsymbol{\Leftarrow}$)}
	Let \Labeling be an \admissibleText{} labeling for $\Instance(\mathcal{A}, \omega)$.
	Recall that by the placement of the sites for the elements in $\mathcal{A}$ and the (sites for the) blocks, the labels for the sites~$s_1$ to $s_N$ can only be inside the $A$-high windows on the boundary.
	We define $\mathcal{A}_1$ as the set of elements whose corresponding site is labeled between~$x_5$ and $b_1$, i.e., in the upper window, and $\mathcal{A}_2$ as the one where the label is between $b_5$ and $y_1$, i.e., in the lower window.
	Observe that while the top- and bottom-most labels of the blocks can slide around freely, we can never enlarge the size of the windows on the boundary beyond~$A$, but possible shrink them.
	However, as the sum of the labels of the sites created for the elements in $\mathcal{A}$ is~$2A$, shrinking the size of a window is never beneficial.
	Hence, each of these windows has only space for that many sites such that the sum of their label heights equals $A$.
	As a consequence, $A = \sum_{a \in \mathcal{A}_1} w(a) = \sum_{a \in \mathcal{A}_2} w(a)$ must hold, i.e., $\left(\mathcal{A}, w\right)$ is a positive instance of \probname{Partition}.
\end{proof}

We would like to point out that the problem \probname{Partition} is only \emph{weakly} \NP-complete.
Therefore, \cref{thm:hardness-1-sided} does not prove that \KSCBLProblemShort{1} is also \NP-hard in the strong sense and such a result would require a reduction from a different problem.
Alternatively, obtaining a pseudo-polynomial algorithm for our problem at hand would show that it is weakly \NP-complete, ruling out strong \NP-hardness under common complexity assumptions.
We refer to the book by Garey and Johnson~\cite{Garey.1979} for an introduction to \NP-completeness and its flavors and leave both directions open for future work.
Despite \cref{thm:hardness-1-sided}, \KSCBLProblemShort{1} is polynomial time solvable for a pre-defined set of fixed \candidatesText{} or uniform-height labels, as we %
see next.

\subsection{Fixed \CandidateText{} \ReferencePointsText{}}
\label{sec:fixed-ports}
In this section, we assume that we are given a set \Candidates of $m \geq n$ \candidatesText{}.
Once \candidateText{} positions for the labels are known, we can compute an ordinary one-sided boundary labeling by employing a dynamic program that evaluates the quality of the leader for the leftmost site on each \candidateText{} and thereby recursively splits the instance into smaller and smaller sub-instances.
However, in the presence of our constraints, not all \candidatesText{} can lead to an \admissibleText{} labeling.
Therefore, when evaluating a \candidateText{}, we have to ensure that no constraint is violated.
Exploiting a connection to the consecutive ones property, we propose a tree-like data structure to efficiently perform these checks.
In the following, we describe this data structure and the dynamic program that uses it in more detail.

\begin{figure}
	\centering
	\includegraphics{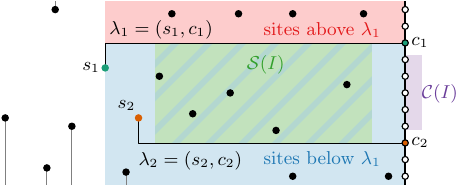}
	\caption{A sub-instance $I = (s_1, c_1, s_2, c_2)$ of our DP-Algorithm and the used notation. Sites outside the sub-instance that might be labeled are indicated with gray leaders.}
	\label{fig:definition-sub-instance}
\end{figure}

We start with noting that Benkert, Haverkort, Kroll, and Nöllenburg~\cite{Benkert.2009} observed that in a planar labeling \Labeling, the leader $\Leader_L$ connecting the leftmost site $s_L \in \Sites$ with some \candidateText{}~$c_L$ splits the instance \Instance into two independent sub-instances, $\Instance_1$ and $\Instance_2$, excluding $s_L$ and~$c_L$.
If we now add to \Instance two artificial sites $s_0$ and $s_{n + 1}$ that are left of all sites in \Instance, and two artificial \candidatesText{} $c_0$ and $c_{m + 1}$, one above \Instance and one below \Instance, we can describe any sub-instance $I$ of \Instance by two leaders $(s_1, c_1)$ and $(s_2, c_2)$ that bound $I$ from above and below, respectively. 
In particular, we can describe \Instance by the
two leaders $(s_0, c_0)$ and $(s_{n + 1}, c_{m + 1})$.
\cref{fig:definition-sub-instance} visualizes this and the following notions.
We refer to the sub-instance as $I = (s_1, c_1, s_2, c_2)$ and let $\Restricted{\Sites}{I}$ and $\Restricted{\Candidates}{I}$ denote the sites and \candidatesText{} in~$I$, \emph{excluding} those used in the definition of $I$, i.e., $\Restricted{\Sites}{I} \coloneqq \{s \in \Sites \mid x(s_1) < x(s),\ x(s_2) < x(s),\ y(c_2) < y(s) < y(c_1)\}$ and $\Restricted{\Candidates}{I} \coloneqq \{c \in \Candidates \mid y(c_2) < y(c) < y(c_1)\}$.
Similarly, for a leader $\Leader = (s, c)$, we say that a site $s'$ with $x(s) < x(s')$ is \emph{above} \Leader if $y(s') > y(c)$ holds and \emph{below} \Leader if $y(s') < y(c)$ holds.

Two more observations about \admissibleText{} labelings can be made:
First, $\Leader_L$ never splits two sites $s, s'\in \Sites$ that are in a grouping constraint $\Group\in \GroupingConstraints$ that does not contain $s_L$, i.e., for which we have¸ $s, s' \in \Group$ and $s_L \notin \Group$.
Second, if $\Leader_L$ %
splits the two sites $s, s' \in \Sites$ with $s$ above $\Leader_L$ and $s'$ below $\Leader_L$, then $s$ is labeled above $s_L$ which itself is labeled above $s'$ in any planar labeling. Thus, to also obtain an \admissibleText{} labeling, this must comply with our ordering constraints~\OrderingConstraints, i.e., we cannot %
have $s' \Order s_L$, $s' \Order s$, or $s_L \Order s$.
Now, we could immediately define a dynamic programming (DP) algorithm that evaluates the induced sub-instances for each leader that adheres to these observations.
However, we would then need to check every constraint in each sub-instance and not make use of implicit constraints given by, for example, overlapping groups.
The following data structure makes these implicit constraints explicit.

\subparagraph{PQ-A-graphs.}
Every labeling \Labeling induces a permutation $\pi$ of the sites by reading the labels from top to bottom. 
Assume that we have at least one grouping constraint, i.e., $k > 0$, and let $M(\Sites, \GroupingConstraints)$ be an $n \times k$ binary matrix with $m_{i,j} = 1$ if and only if $s_i \in \Group_j$.
We call $M(\Sites, \GroupingConstraints)$ the \emph{sites vs.~groups} matrix, and observe that \Labeling satisfies the constraint $\Group_j$ if and only if the ones in the column~$j$ of $M(\Sites, \GroupingConstraints)$ are consecutive after we order the rows of $M(\Sites, \GroupingConstraints)$ according to~$\pi$.
If a permutation $\pi$ exists such that this holds (simultaneously) for all columns of $M(\Sites, \GroupingConstraints)$, then it has the so-called \emph{consecutive ones property} (\emph{C1P})~\cite{Fulkerson.1965}.
This leads to the following observation.
\begin{observation}
	\label{lem:grouping-constraints-consecutive-ones}
	\GroupingConstraints are consistent for \Sites if and only if $M(\Sites, \GroupingConstraints)$ has the C1P.
\end{observation}

Booth and Lueker~\cite{Booth.1976} proposed an algorithm to check whether a binary matrix has the C1P. Their algorithm uses a PQ-tree to keep track of the allowed row permutations.
A \emph{PQ-tree} $\tau$, for a given set $\mathcal{A}$ of elements, is a rooted tree with one leaf for each element of $\mathcal{A}$ and two different types of internal nodes~$t$: P-nodes, where we can freely permute the children of $t$, and Q-nodes, where the children of $t$ can only be inversed~\cite{Booth.1976}.
\cref{lem:grouping-constraints-consecutive-ones} tells us that every family of consistent grouping constraints \GroupingConstraints{} can be represented by a PQ-tree~$\tau$, i.e., $\tau$ admits the permutation $\pi$ of $\Leaves{\tau}$ if and only if the labels, if ordered from top to bottom according to $\pi$, induce a (not necessarily planar) labeling~\Labeling that respects \GroupingConstraints.
Note that any PQ-tree, apart from the empty PQ-tree, admits at least one permutation of its leaves.
Therefore, inconsistent grouping constraints cannot be represented by a (non-empty) PQ-tree.
Note that while we can represent a family of consistent grouping constraints \GroupingConstraints by a PQ-tree $\tau$, it is in general not true that for every grouping constraint $\Group \in \GroupingConstraints$ we have a node $t$ in $\tau$ whose leaves contain exactly the elements of \Group.
However, in the special case of non-overlapping grouping constraints, each grouping constraint indeed corresponds to a P-node in $\tau$. 
In either case, we can interpret any node $t$ of $\tau$ as a grouping constraint and we call these the \emph{canonical groups} of $\tau$.
\begin{figure}
	\centering
	\includegraphics{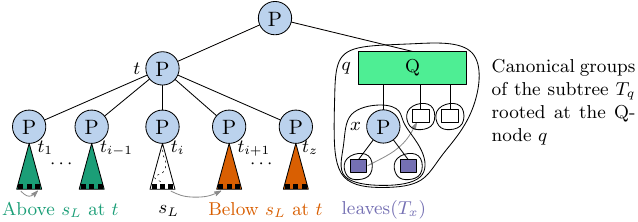}
	\caption{A sample PQ-A-\NewText{g}raph together with the used terminology. Leaves are indicated by squares and ordering constraints by the arrows.}
	\label{fig:dp-pqa}
\end{figure}

While \cref{lem:grouping-constraints-consecutive-ones} implies that PQ-trees can represent families of consistent grouping constraints, it is folklore that directed acyclic graphs can be used to represent partial orders, i.e., our ordering constraints.
We now combine these two data structures into \emph{PQ-A-graphs}.
\begin{definition}[PQ-A-graph]
	\label{def:pq-a-graph}
	Let \Sites be a set of sites, \GroupingConstraints be a family of consistent grouping constraints, and \OrderingConstraints be a partial order on \Sites.
	The \emph{PQ-A-graph} $\PQAGraph = (\PQTree{}, A)$ consists of the PQ-tree \PQTree{} that represents \GroupingConstraints, on whose leaves we embed the arcs $A$ of a directed graph representing \OrderingConstraints.
\end{definition}
We let $T_i$ denote the \emph{subtree} in the underlying PQ-tree \PQTree{} rooted at the node~$t_i$ and %
\Leaves{T_i} the leaf set of $T_i$.
\cref{fig:dp-pqa} shows a PQ-A-graph and the introduced terminology.
Furthermore, observe that checking on the consistency of $\ConstraintsLong$ is equivalent to solving the \probname{Reorder} problem on \PQTree{} and~\OrderingConstraints, i.e., asking whether we can re-order \Leaves{\PQTree{}} such that the order induced by reading them from left to right extends the partial order~\OrderingConstraints~\cite{Klavik.2017}.
\begin{restatable}{lemma}{pqaGraphProbertiesLemma}
    \label{lem:pqa-graph-properties}
    Let \Sites be a set of $n$ sites, \GroupingConstraints be $k$ grouping constraints and \OrderingConstraints be $r$ ordering constraints. 
    We can check whether $\ConstraintsLong$ are consistent for \Sites and, if so, create the PQ-A-graph \PQAGraph in \BigO{n + k + r + \sum_{\Group \in \GroupingConstraints} \Size{\Group}} time.
    \PQAGraph uses \BigO{n + r} space.
\end{restatable}
\begin{proof}
	We prove the statements of the lemma individually.
	
	\proofsubparagraph*{Running Time of the Consistency Checks.}
	Regarding the grouping constraints, we discussed in \cref{lem:grouping-constraints-consecutive-ones} that \GroupingConstraints being consistent for~\Sites is equivalent to the sites vs.~groups matrix $M(\Sites, \GroupingConstraints)$ having the C1P.
	Booth and Lueker~\cite{Booth.1976} propose an algorithm to check whether a binary matrix $M$ has the C1P.
	Since their algorithm decomposes $M$ into its columns to build a PQ-tree~\PQTree{} out of those, which for~$M(\Sites, \GroupingConstraints)$ corresponds to the groups $\Group \in \GroupingConstraints$, we do not need to compute $M(\Sites, \GroupingConstraints)$ but can directly work with \GroupingConstraints.
	Their algorithm has a running time of \BigO{n + k + \sum_{\Group \in \GroupingConstraints} \Size{\Group}}~\cite[Theorem~6]{Booth.1976}.
	If it outputs that $M(\Sites, \GroupingConstraints)$ does not have the C1P, then we know that the constraints \ConstraintsLong cannot be consistent for \Sites, as \GroupingConstraints is not consistent for \Sites.
	On the other hand, if it outputs that $M(\Sites, \GroupingConstraints)$ has the C1P, we can modify the algorithm to also return the PQ-tree~\PQTree{}, that the algorithm internally maintains, without spending additional time.
	Regarding the ordering constraints, it remains to check whether \PQTree{} allows for a permutation that extends the partial order \OrderingConstraints on the sites.
	This is equivalent to the instance $(\PQTree{}, \OrderingConstraints)$ of the \probname{Reorder} problem, which can be solved in \BigO{n + r} time~\cite[Proposition 2.4]{Klavik.2017}.
	
	The claimed running time for checking the consistency of $\ConstraintsLong$ for \Sites follows then readily.
	For the rest of the proof, we assume that the constraints \ConstraintsLong are consistent, as otherwise the corresponding PQ-A-graph \PQAGraph might not be defined.
	
	\proofsubparagraph*{Creation Time of $\boldsymbol{\PQAGraph}$.}
	We have already concluded that we can obtain the PQ-tree \PQTree{} in \BigO{n + k + \sum_{\Group \in \GroupingConstraints} \Size{\Group}} time using the algorithm by Booth and Lueker~\cite{Booth.1976}.	
	In the following, we assume that we maintain a look-up table that returns for each site $s \in \Sites$ the corresponding leaf in \PQTree{}.
	Since we never add or remove a leaf, maintaining this table does not increase the asymptotic running time of creating the PQ-tree \PQTree{}.
	
	To finish the creation of \PQAGraph, we have to enrich \PQTree{} by the ordering constraints~\OrderingConstraints.
	Let $s \Order s'$ be one of those constraints.
	Since we can find the leaves for $s$ and $s'$ in \PQTree{} in constant time using our look-up table, adding the corresponding arc to~\PQTree{} takes \BigO{1} time.
	This sums up to~\BigO{r} and together with above arguments we get \BigO{n + k + r + \sum_{\Group \in \GroupingConstraints} \Size{\Group}}.   
	
	\proofsubparagraph*{Space Consumption of $\boldsymbol{\PQAGraph}$.}
	Regarding the space consumption of the underlying PQ-tree \PQTree{}, we first note that Booth and Lueker assumed that they work with \emph{proper} PQ-trees.
	This means that any P-node has at least two and any Q-node at least three children, respectively, i.e., there are no (chains of) nodes with a single child~\cite{Booth.1976}.
	Hence, \PQTree{} uses \BigO{n} space~\cite{Jiang.2020}.
	To embed the arcs on the leaves of \PQTree{}, we can use adjacency sets, i.e., adjacency lists where we use sets to store the neighbors of the nodes.
	This gives us constant time look-up and \BigO{n + r} space, and combined with the space consumption of the PQ-tree, we get \BigO{n + r}.
\end{proof}

\subparagraph{The Dynamic Programming Algorithm.}
Let $I = (s_1, c_1, s_2, c_2)$ be a sub-instance and $s_L$ the leftmost site in \Restricted{\Sites}{$I$}.
Furthermore, let $\PQAGraph(s_1, s_2)$ denote the subgraph of the PQ-A-graph~\PQAGraph rooted at the lowest common ancestor of~$s_1$ and $s_2$ in \PQAGraph. 
Note that $\PQAGraph(s_1, s_2)$ contains at least the sites in $\Restricted{\Sites}{I}$, together with~$s_1$ and $s_2$.
Therefore, it represents all constraints relevant for the sub-instance~$I$.
Other constraints either do not affect sites in $I$, i.e., are represented by other parts of \PQAGraph, or are trivially satisfied.
In particular, observe that all constraints induced by nodes further up in \PQAGraph affect a super-set of $\Restricted{\Sites}{I}$ and, therefore, %
we check them in some sub-instance $I'$ that contains~$I$, i.e., that contains $s_1$, $s_2$, $c_1$, and~$c_2$. 
Now imagine that we want to place the label~$\Label_L$ for~$s_L$ at the \candidateText{}{} $c_L \in \Restricted{\Candidates}{I}$.
Let \Admissible{I, \PQAGraph, c_L} be a procedure that checks the following criteria.
\begin{enumerate}
	\item The label $\Label_L$ does not overlap with the labels $\Label_1$, placed at $c_1$, and $\Label_2$, placed at $c_2$, for the sites $s_1$ and $s_2$, respectively, that define the sub-instance $I$.
	\item The leader $\Leader_L$ does not intersect with a site $s' \in \Restricted{\Sites}{I}$, $s' \neq s_L$.
	\item In the resulting sub-instances $I_1 = (s_1, c_1, s_L, c_L)$ and $I_2 = (s_L, c_L, s_2, c_2)$ are enough \candidatesText{} for all sites, i.e., $\Size{\Restricted{\Sites}{I_i}} \leq \Size{\Restricted{\Candidates}{I_i}}$, for $i = 1,2$.
	\item Placing the label $\Label_L$ at $c_L$ and thus having the leader $\Leader_L = (s_L, c_L)$ respects the constraints expressed by $\PQAGraph(s_1, s_2)$.
\end{enumerate}
The first three criteria ensure planarity with respect to the already fixed labeling and that the resulting sub-instances contain enough \candidatesText{} for the respective sites.
The last criterion guarantees that the labeling eventually %
respects the constraints (expressed by $\PQAGraph(s_1, s_2)$).
To perform the checks for that criterion efficiently, we make use of the procedure \RespectsConstraints{I, \PQAGraph, \Leader_L} defined as follows. 

Consider again \cref{fig:dp-pqa} for the following description.
Let $t_L$ be the leaf for $s_L$ in $\PQAGraph(s_1, s_2)$.
There is a unique path from $t_L$ to \rootOfTree{\PQAGraph(s_1, s_2)}, which we traverse bottom up and consider each internal node on it.
Let $t$ be such a node with the children $t_1, \dots, t_z$ in this order from left to right.
Let~$T_i$, $1 \leq i \leq z$, be the subtree that contains the site $s_L$, rooted at $t_i$.
The labels for all sites represented by $\Leaves{T_1}, \dots, \Leaves{T_{i - 1}}$ %
are placed above $\Label_L$ in any labeling \Labeling of \Sites in which the children of $t$ are ordered as stated.
Therefore, we call these sites \emph{above~$s_L$} (at~$t$).
Analogously, the sites represented by $\Leaves{T_{i + 1}}, \dots, \Leaves{T_z}$ are \emph{below $s_L$} (at~$t$).
The sites represented by \Leaves{T_i} are neither above nor below $s_L$ at $t$.
It is important to note that we use two different notions of \emph{above}/\emph{below}.
On the one hand, sites can be above a node $t$ in the PQ-A-graph $\PQAGraph(s_1, s_2)$, which depends on the order of the children of~$t$.
On the other hand, a site can also be above a leader \Leader, which depends on the (geometric) position of \Leader and is independent of $\PQAGraph(s_1, s_2)$.
Recall \cref{fig:dp-pqa} and compare it with \cref{fig:definition-sub-instance} for the former and latter notion of above and below, respectively.

If $t$ is a P-node, we seek a permutation $\pi$ of the children $t_1, \dots, t_z$ of $t$ in which all the sites in \Restricted{\Sites}{$I$} above $s_L$ at $t$ (in the permutation $\pi$) are above $\Leader_L$, and all the sites in \Restricted{\Sites}{$I$} below $s_L$ at $t$ (in the permutation $\pi$) are below $\Leader_L$.
This means that it cannot be the case that some sites of the same subtree $T$ that does not contain $s_L$ are above $\Leader_L$, while others are below $\Leader_L$, as this implies that we violate the canonical grouping constraint induced by~$T$.
To not iterate through all possible permutations, we distribute the children of $t$, except $t_i$, into two sets, $C_{\text{above}}$ and $C_{\text{below}}$, depending on whether the sites they represent should be above or below $s_L$ at $t$.

\begin{figure}
	\centering
	\begin{subfigure}[t]{.45\linewidth}
		\centering
		\includegraphics[page=2]{figure_06}
		\subcaption{The PQ-A-graph $\PQAGraph(s_1, s_2)$ with $s_1 \in \Leaves{T_i}$.}
		\label{fig:dp-1-sided-permutation-s-s1_a}
	\end{subfigure}
	\quad
	\begin{subfigure}[t]{.45\linewidth}
		\centering
		\includegraphics[page=1]{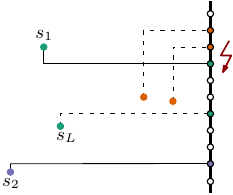}
		\subcaption{A (sub-)instance $I$, where the wrong permutation of the children of $t$ in $\PQAGraph(s_1, s_2)$ from~(\subref{fig:dp-1-sided-permutation-s-s1_a}) would label the orange sites outside~$I$.}
		\label{fig:dp-1-sided-permutation-s-s1_b}
	\end{subfigure} 
	\caption{In this situation, $C_{\text{above}}$ must not contain subtrees with sites from the sub-instance.}
	\label{fig:dp-1-sided-permutation-s-s1}
\end{figure}

More concretely, we check if all sites from $\Leaves{T_j} \cap \Restricted{\Sites}{I}$ are above or below the leader $\Leader_L$ and assign $t_j$ to $C_{\text{above}}$ or $C_{\text{below}}$, respectively.
Furthermore, if $\Leaves{T_j}$ only contains sites outside~$I$, then the leaders from $s_1$ or $s_2$ separate the sites from $T_j$ and $\Restricted{\Sites}{I}$.
As these sites are part of an instance $I'$ that contains~$I$, for which we separately ensure that constraints relating a site from~$T_j$ and one from $\Restricted{\Sites}{I}$ are respected, we can ignore $t_j$ in this case.
However, for other cases we also have to ensure that this assignment is consistent with $I$,~$s_1$, and $s_2$.
In particular, if $s_1 \in \Leaves{T_j}$, then $T_j$ can contain some sites in $I$ and other sites outside $I$.
As $s_1$ is above~$s_L$ at~$t$ by the definition of $I$, we must put $t_j$ in $C_{\text{above}}$.
Hence, we check whether all sites in $\Leaves{T_j} \cap \Restricted{\Sites}{I}$ are above $\Leader_L$.
Furthermore, if $s_1 \in \Leaves{T_i}$ as in \cref{fig:dp-1-sided-permutation-s-s1}, then $C_{\text{above}}$ must not contain a child $t_j$ of $t$, as the sites in $T_j$ would then be labeled outside $I$, violating the definition of $I$.
The case where $s_2$ is in one of those subtrees is handled analogously.

The checks that have to be performed if $t$ is a Q-node are conceptually the same, but simpler, since Q-nodes only allow to inverse the order:
Either all sites above $s_L$ at $t$ are above~$\Leader_L$, and all sites below $s_L$ at $t$ are below~$\Leader_L$.
Or all sites above $s_L$ at $t$ are below~$\Leader_L$, and all sites below $s_L$ at $t$ are above $\Leader_L$.
In the former case, we keep the order of the children at the node $t$ as they are.
In the latter case, we inverse the order of the children at the node~$t$.
Note that if a child of $t$ contains the sites $s_1$, $s_2$, or sites outside the sub-instance $I$ but in $\PQAGraph(s_1, s_2)$, one of the two allowed inversions is enforced by the definition of~$I$, i.e., we must pick the order that places $s_1$ before the sites from $\Restricted{\Sites}{I}$ and $s_2$ after them.
Of course, we still need to ensure that this order is consistent with the position of the sites with respect to the leaders for $s_L$ and $s_1$ or $s_2$.

Until now we only verified that we adhere to the grouping constraints.
To ensure that we do not violate an ordering constraint, we maintain a look-up table that stores for each site whether it belongs to $C_{\text{above}}$, $C_{\text{below}}$, or $T_i$.
We use this to check in constant time whether an ordering constraint is violated or not.
In particular, an ordering constraint is violated if the corresponding arc runs from $C_{\text{below}}$ to $T_i$ or to $C_{\text{above}}$, or from~$T_i$ to $C_{\text{above}}$.

We now bound the (overall) running time of these checks and make the following observations.

First, the length of the path from $t_L$ to \rootOfTree{\PQAGraph(s_1, s_2)} is bounded by the height of $\PQAGraph(s_1, s_2)$ which is in \BigO{n}. 
Second, we query the position for each site $s \in \Sites \setminus \{s_1, s_2\}$ at most once to determine its position with respect to $\Leader_L$ and/or check whether it is in the sub-instance.
Afterwards, i.e., when we proceed to the next node on the path towards \rootOfTree{\PQAGraph(s_1, s_2)}, $s \in \Leaves{T_i}$ holds.
Since we only query membership of $s_1$ and/or $s_2$ in $\Leaves{T_i}$, it follows that we do not consider $s$ anymore.
Hence, these checks take for the node $s$ overall $\BigO{1}$ time and for the nodes $s_1$ and $s_2$ \BigO{n} time, since we check membership in $\Leaves{T_i}$ on every of the \BigO{n} nodes on the path.
Overall, we have a(n amortized) running time of \BigO{n} to check grouping constraint.
Third, observe that we also check every ordering constraint at most once.
Note that we have to check on every of the \BigO{n} nodes on the path that there is no site in $T_i$ that has a violated ordering constraint with a site in $C_{\text{above}}$ or $C_{\text{below}}$.
However, we can, for a node $t$ on the path, iterate through all sites in $C_{\text{above}}$ and $C_{\text{below}}$, and check for each of their incident arcs whether they represent a violated ordering constraint.
Since these sites are for upcoming nodes on the path towards \rootOfTree{\PQAGraph(s_1, s_2)} in $\Leaves{T_i}$, we query these arcs only once.
By doing so, we check each ordering constraint at most twice, since each of its involved sites can be in $C_{\text{above}}$ or $C_{\text{below}}$ at most once.
In particular, this avoids checking all ordering constraints for every site in $\Leaves{T_i}$, as this could lead to checking the same ordering constraint \BigO{n} times.
As each ordering constraint can be checked in \BigO{1} time, we obtain an overall running time of \BigO{r} to check all ordering constraints.

We say that $c_L$ \emph{respects the constraints for $s_L$} imposed by $\PQAGraph(s_1, s_2)$ in the sub-instance $I = (s_1, c_1, s_2, c_2)$ if it respects them at every node $t$ on the path from $s_L$ to the root of $\PQAGraph(s_1, s_2)$.
\RespectsConstraints{I, \PQAGraph, \Leader_L} performs these checks for each node on the path from $s_L$ to \rootOfTree{\PQAGraph(s_1, s_2)}. %
Using the arguments from above, we conclude that \RespectsConstraints{I, \PQAGraph, \Leader_L} runs in $\BigO{n + r}$ time.
By the same arguments, this also captures the computation of the look-up tables to check for violated ordering constraints.
In the following lemma, we show that \Admissible{I, \PQAGraph, c_L}
takes \BigO{n + r + \log m} time.
\begin{restatable}{lemma}{feasibleRunningTimeLemma}
    \label{lem:feasible-running-time}
	Let $I = (s_1, c_1, s_2, c_2)$ be a sub-instance of our DP-Algorithm with the constraints expressed by a PQ-A-graph \PQAGraph.
	We can check whether the \candidateText{} $c_L \in \Restricted{\Candidates}{I}$ is \admissibleText{} for the leftmost site $s_L \in \Restricted{\Sites}{I}$ in \BigO{n + r + \log m} time using \Admissible{I, \PQAGraph, c_L}, where~$n = \Size{\Sites}$, $m = \Size{\Candidates}$, and $r$ is the number of ordering constraints.
\end{restatable}
\begin{proof}
	Checking admissibility of a \candidateText{} $c_L \in \Restricted{\Candidates}{I}$ for the leftmost site $s_L \in \Restricted{\Sites}{I}$ consists of checking whether the following four criteria are satisfied.
	
	\proofsubparagraph*{Criterion~1.}
	In Criterion~1, we must ensure that the label $\Label_L$ for the leftmost site does not overlap already placed labels.
	As we have fixed \candidatesText{} $c_1$, $c_2$, and $c_L$, we can do this in constant time as we only have to check whether $y(c_2) + (h(\Label_2) + h(\Label_L))/ 2 \leq y(c_L) \leq y(c_1) - (h(\Label_1) + h(\Label_L)) / 2$ holds.
	
	\proofsubparagraph*{Criterion~2.}
	For Criterion~2, we have to ensure that no leader-site crossing is introduced. 
	Hence, we check for each site $s' \in \Restricted{\Sites}{I}$ with $x(s_L) < x(s')$ that it does not have the same $y$-coordinate as the \candidateText{} $c_L$, i.e, it cannot hold $y(s') = y(c_L)$.
	This takes \BigO{n} time.
	
	\proofsubparagraph*{Criterion~3.}
	The third criterion states that the resulting sub-instances must contain at least as many \candidatesText{} as sites.
	To check this criterion efficiently, we assume that we can access a range tree storing the sites, and a list containing the \candidatesText{} sorted by their $y$-coordinate.
	We %
	account for this when we discuss the overall properties of our DP-Algorithm.
	With this assumption, we can compute the number of \candidatesText{} in a sub-instance by running two binary searches for $c_1$ and $c_2$ which takes \BigO{\log m} time.
	To count the number of sites in the sub-instance, we can run a counting range query on the sites, which takes \BigO{\log n} time~\cite{Berg.2008}.
	We arrive at a running time of \BigO{\log n + \log m}.
	
	\proofsubparagraph*{Criterion~4.}
	The last criterion states that we must respect the constraints expressed by $\PQAGraph(s_1, s_2)$.
	Observe that we do not need to compute $\PQAGraph(s_1, s_2)$ as we can simply traverse the path from the leaf for the leftmost site $s_L~\in~\Restricted{\Sites}{I}$, to the root of \PQAGraph and stop once we reach the root of $\PQAGraph(s_1, s_2)$.
	This node is the least common ancestor of $s_1$ and $s_2$. Hence, we reach the root of $\PQAGraph(s_1, s_2)$ once we have $s_1$ \emph{and} $s_2$ for the first time in the subtree.
	The time required to check Criterion~4 coincides with the running time of \RespectsConstraints{I, \PQAGraph, \Leader_L}, which we have upper-bounded by \BigO{n + r}.
	
	Combining all, we get a running time of \BigO{n + r + \log m}.
\end{proof}

For a sub-instance $I = (s_1, c_1, s_2, c_2)$, we store in a table $D$ the value $f(\Labeling^*)$ of an optimal \admissibleText{} labeling $\Labeling^*$ on $I$ or $\infty$ if none exists.
If $I$ does not contain a site we set $D[I] = 0$.
Otherwise, we use the following relation, where the minimum of the empty set is $\infty$.
\begin{equation*}
	D[I] = \min_{{\substack{c_L \in \Restricted{\Candidates}{I}\ \text{where}\\\ \Admissible{I, \PQAGraph, c_L}\\\ \text{is true}}}} \left(D[(s_1, c_1, s_L, c_L)] + D[(s_L, c_L, s_2, c_2)]\right) + f((s_L, c_L))
\end{equation*}
To show correctness of our DP-Algorithm, one can use a proof analogous to the one of
Benkert et al.~\cite{Benkert.2009}, who %
propose a similar dynamic program to compute a one-sided labeling with \Po-leaders and a similar-structured optimization function, combined with the fact that we consider only those \candidatesText{} that are \admissibleText{} for $s_L$.
Recall that, after adding the
artificial sites $s_0$ and $s_{n + 1}$, and \candidatesText{} $c_0$ and $c_{m + 1}$, %
we can describe any sub-instance by a tuple $I= (s_1, c_1, s_2, c_2)$, and in particular the sub-instance for \Instance by $I_0 = (s_0, c_0, s_{n + 1}, c_{m + 1})$.
Hence, $D[I_0]$ %
stores in the end $f(\Labeling^*)$, or $\infty$, if \Instance does not possess an \admissibleText{} labeling.
If the labeling $\Labeling^*$ exists, it can be obtained using standard techniques.
As two sites and two \candidatesText{} describe a sub-instance, there are up to \BigO{n^2m^2} possible sub-instances to evaluate.
We then fill the table~$D$ top-down using memoization.
This guarantees us that we have to evaluate each sub-instance~$I$ at most once, and only those that arise from \admissibleText{} \candidatesText{}.
The running time of evaluating a single sub-instance is dominated by the time required to determine for each \candidateText{} whether it is \admissibleText{}.
Combined with the size of the table $D$, we get the following.
\begin{restatable}{theorem}{dpPropertiesTheorem}
    \label{thm:dp-running-time-space} 
    \KSCBLProblemShort{1} for $n$ sites, $m$ \emph{fixed} \candidatesText{}, $r$ ordering, and $k$ grouping constraints \GroupingConstraints can be solved in \BigO{n^4m^3\log m + k + \sum_{\Group \in \GroupingConstraints} \Size{\Group}} time and \BigO{n^2m^2} space.
\end{restatable}
\begin{proof}
	Let \Instance be an instance of \KSCBLProblemShort{1} with the constraints $\ConstraintsLong$.
	From \cref{lem:pqa-graph-properties}, we know that we can check in \BigO{n + k + r + \sum_{\Group \in \GroupingConstraints} \Size{\Group}} time whether the constraints \ConstraintsLong are consistent for \Sites.
	Let us assume that they are, as otherwise \Instance does not possess an \admissibleText{} labeling \Labeling.
	Therefore, we can obtain, in this time, the corresponding PQ-A-graph \PQAGraph that uses \BigO{n + r} space.
	
	As further preprocessing steps, we create an additional list storing the \candidatesText{} $c \in \Candidates$ sorted by $y(c)$ in ascending order and a range tree on the sites.
	We can do the former in \BigO{m \log m} time and \BigO{m} space, and the latter in \BigO{n \log n} time and space~\cite{Berg.2008}.
	In addition, we compute for each internal node~$t$ of the PQ-A-graph \PQAGraph the canonical group induced by the subtree rooted at~$t$. %
	This can be done %
	in \BigO{n^2} time and space by traversing \PQAGraph bottom-up, as~\PQAGraph has \BigO{n} internal nodes and each canonical group is of size at most~$n$.
	Furthermore, we create a look-up table for $f((s, c))$, i.e., we compute $f((s, c))$ for all $(s,c) \in \Sites \times \Candidates$.
	Note that $\Sites \times \Candidates = \Lambda$ for fixed \candidatesText{} (and fixed \portsText{}).
	Creating this look-up table consumes \BigO{nm} space and requires \BigO{nm} time, assuming that $f(\cdot)$ can be evaluated in constant time.
	Note that this is true if we seek an \admissibleText{}, length-, or bend-minimal labeling.
	
	We then continue filling the DP-table $D$ top-down using memoization.
	This guarantees that we have to evaluate each sub-instance $I$ at most once, and only those sub-instances that arise from \admissibleText{} (candidate) leaders.
	For each such $I$, we have to, when evaluating the recurrence relation, check for \BigO{m} \candidatesText{} if they are \admissibleText{} for the leftmost site $s_L$.
	We can obtain $s_L$ in \BigO{n} time.
	For each \candidateText{} $c_L$, we first have to check whether it is \admissibleText{} for $s_L$, which we can do in \BigO{n + r + \log m} time due to \cref{lem:feasible-running-time}.
	Since we have pre-computed all values for $f(\cdot)$, this is the overall time required for a single \candidateText{}~$c_L$.
	Hence, evaluating all \candidatesText{} takes $\BigO{m \left(n + r + \log m\right)}$ time.
	As there are \BigO{n^2m^2} possible sub-instances~$I$ that we have to evaluate in the worst case, our DP-Algorithm solves \KSCBLProblemShort{1}, for a given instance \Instance, in $\BigO{n^2m^2(n + m \left(n + r + \log m\right)) + k + \sum_{\Group \in \GroupingConstraints} \Size{\Group}} = \BigO{n^3m^3 + n^2m^3r + n^2m^3\log m + k + \sum_{\Group \in \GroupingConstraints} \Size{\Group}}$ time using \BigO{n^2m^2} space.
	Note that the above bounds dominate the time and space required for the remaining preprocessing steps.
	Observe that $r = \BigO{n^2}$ holds.
	Hence, if~$r$ is small, i.e., $r~=~\BigO{n}$, above bounds can be simplified to $\BigO{n^3m^3 + n^2m^3\log m + k + \sum_{\Group \in \GroupingConstraints} \Size{\Group}}$.
	On the other hand, if $r$ is large, i.e., $r = \Theta(n^2)$, above bounds yield $\BigO{n^4m^3 + n^2m^3\log m  + k + \sum_{\Group \in \GroupingConstraints} \Size{\Group}}$. 
	Although yielding a higher bound, to ease readability, we %
	upper-bound the latter running time by $\BigO{n^4m^3\log m  + k + \sum_{\Group \in \GroupingConstraints} \Size{\Group}}$, resulting in the claimed bounds.
\end{proof}
Real-world instances often consist of less than 50 sites~\cite{Niedermann.2017} and we do not expect the number of \candidatesText{} to be significantly larger than the number of sites.
Hence, the running time of \cref{thm:dp-running-time-space} does not immediately rule out the practical applicability of our results.
Indeed, initial experiments for uniform-height labels confirmed that our algorithm terminates within a few seconds for instances with realistic sizes of up to 25 sites and 50 \candidatesText{}~\cite{Barth.2019,Gedicke.2023}; see \cref{sec:experiments} for details and \cref{fig:illustration-motivation_b} for an example.
In fact, dynamic programming is frequently used to obtain exact polynomial-time algorithms in external labeling~\cite{Bekos.2021} and it is not uncommon that such algorithms have high running times of up to \BigO{n^6} and \BigO{n^9} for the one- and two-sided setting with \Po-leaders, respectively~\cite{Benkert.2009,Fink.2016,Kindermann.2016,Klawitter.2023}.
Finally, we observed in our experiments that the position of the \candidatesText{} influenced the feasibility of an instance, which makes considering sliding \candidatesText{} interesting and relevant.

\subsection{Sliding \CandidateText{} \ReferencePointsText{} with Uniform-Height Labels}
\label{sec:sliding-ports}
\begin{figure}[t]
	\centering
	\begin{subfigure}[t]{.45\linewidth}
		\centering
		\includegraphics[page=1]{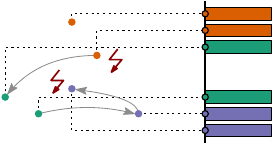}
		\subcaption{We cannot avoid crossings if we want to respect the constraints.}
		\label{fig:sliding-labels-motivation_a}
	\end{subfigure}
	\quad
	\begin{subfigure}[t]{.45\linewidth}
		\centering
		\includegraphics[page=2]{figure_07}
		\subcaption{
			An alternative set of \candidatesText{} to~(\subref{fig:sliding-labels-motivation_a}) which allows for an \admissibleText{} labeling.}
		\label{fig:sliding-labels-motivation_b}
	\end{subfigure} 
	\caption{
		An instance whose admissibility depends on the position of the \candidatesText{}.
	}
	\label{fig:sliding-labels-motivation}
\end{figure}

\looseness=-1
\cref{fig:sliding-labels-motivation} underlines that fixed \candidatesText{} have the limitation that the admissibility of an instance depends on the choice and position of the \candidatesText{}. %

By allowing the labels to slide along a sufficiently long vertical boundary line, we remove this limitation.
To avoid the \NP-hardness shown in \cref{sec:one-sided-hardness}, we require that all labels now have a uniform height of some $h > 0$.

\begin{figure}
	\centering
	\begin{subfigure}[t]{.45\linewidth}
		\centering
		\includegraphics[page=1]{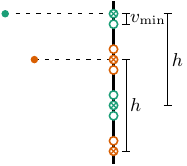}
		\subcaption{Induced \candidatesText{} as in~\cite{Fink.2016} (marked by a cross) and the extension to canonical \candidatesText{}.}
		\label{fig:canonical-ports_a}
	\end{subfigure}
	\quad
	\begin{subfigure}[t]{.45\linewidth}
		\centering
		\includegraphics[page=1]{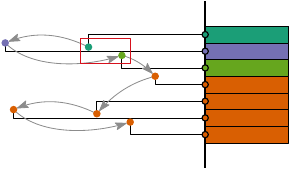}
		\subcaption{
			In the unique labeling for this instance, we can force the leader for the purple label to pass arbitrarily close between the green sites.}
		\label{fig:canonical-ports_b}
	\end{subfigure} 
	\caption{The set of \candidatesText we construct and~(\subref{fig:canonical-ports_b}) %
		an example where we need to extend the induced \candidatesText{} by Fink and Suri~\cite{Fink.2016}.}
	\label{fig:canonical-ports}
\end{figure}

In this section, we show that we can re-obtain tractability of the problem by defining for each site $s$ a set of $\BigO{n}$ \candidatesText{}.
We would like to re-use an idea of Fink and Suri~\cite{Fink.2016} who observe for a similar labeling problem that, if there is an optimal crossing free solution, then there is an optimal crossing free solution in which all labels are placed on candidates from a restricted set of candidates of quadratic size, which contains the horizontal projections of all sites and copies at regular $y$-offsets, shown in \Cref{fig:canonical-ports_a} with the crossed candidates. 
However, constraints make instances with sliding candidates surprisingly challenging. 
In particular, there are instances, such as the one in \Cref{fig:canonical-ports_b}, that admit an admissible labeling but not on the above-mentioned candidates.
The instance from \Cref{fig:canonical-ports_b} furthermore underlines that it is in general
not possible to define the set of candidates purely on the position of individual sites without assuming further restrictions.
In particular, the labeling from \Cref{fig:canonical-ports_b} is the only admissible labeling apart from trivial label movements.
We observe that the leader for the purple site must pass through the two green sites marked by a red box. Since we can place them arbitrarily close to the (hypothetical position of the) leader, we can effectively encapsulate the corresponding label anywhere on the boundary, (almost) unrelated to the position of its site.

A further challenge arises once we are not only interested in some \admissibleText{} labeling, but one that admits short leaders.
Consider the instance from \cref{fig:sliding-labels-motivation_b}, which admits, even on sliding \candidatesText{}, an \admissibleText{} but no length-minimal \admissibleText{} labeling.
The reason for the absence of the latter labeling is the red leader, which we can always move by a small $\varepsilon > 0$ closer to the purple site marked with a red box, which reduces the length of the leader by $\varepsilon$.

From an algorithmic perspective, it is desirable that instances with some \admissibleText{} labeling also admit one that is optimal.
Therefore, to ensure the existence of length-minimal labelings, we enforce from now on that for each site $s$ its leader $\Leader_s$ maintains a minimal vertical distance of $0 < v_{\min} < h$ to each non-incident site $s' \neq s$ with $x(s) < x(s')$.
Observe that %
this requirement does not enforce the vertical length of $\Leader_s$ to be at least $v_{\min}$ and thus still allows for a labeling where some leaders have no bends.
Furthermore, observe that this renders depending on the choice of $v_{\min}$ some instances, such as the one from \cref{fig:canonical-ports_b} infeasible.
However, we do not see this as a limitation because we can, for example, set $v_{\min}$ to the smallest distance that can still be displayed.
After this point, label movements would in either case no longer make a perceivable difference.
Furthermore, note that analogous requirements are already often expressed for real-world labelings~\cite{Niedermann.2017}.
Finally, we can readily patch this requirement into our DP-Algorithm from
\cref{sec:fixed-ports}.
This does not increase its running time, but requires that also leaders in instances with fixed \candidatesText{} maintain said distance.

We now define, based on the position of the sites and $v_{\min}$, a set of \BigO{n^2} \candidatesText{} such that there exists an \admissibleText{} labeling on these (fixed) \candidatesText{}, if the instance with sliding \candidatesText{} possesses an \admissibleText{} labeling.
For each site $s \in \Sites$, we define the 
following set of \candidatesText{}.
\begin{align*}
	\Candidates(s) \coloneqq \{y(s) + jh,\ y(s) + jh \pm v_{\min},\ y(s) - jh,\ y(s) - jh \pm v_{\min} \mid 0 \leq j \leq n\}
\end{align*}
We define the set of \emph{canonical \candidatesText{}} $\Candidates(\Sites)$ as $\Candidates(\Sites) \coloneqq \bigcup_{s\in \Sites} \Candidates(s)$, some of which are depicted in \cref{fig:canonical-ports_a}, and show the following.
\begin{lemma}
	\label{lem:sliding-ports-discretization-sufficient}
	Let \Instance be an instance of \KSCBLProblemShort{1} with uniform-height labels and sliding \candidatesText{}.
	If \Instance possesses an \admissibleText{} labeling, it also has one with \candidatesText{} from $\Candidates(\Sites)$.
\end{lemma}
\begin{proof}
	Our proof builds on arguments used by Fink and Suri for a similar result~\cite[Lemma~1]{Fink.2016}.
	Furthermore, we %
	call a maximal set of touching (but non-overlapping) labels a \emph{stack}, following nomenclature used by N\"ollenburg, Polishchuk, and Sysikaski~\cite{Noellenburg.2010}.
	The overall approach is to transform an \admissibleText{} labeling \Labeling into a labeling~$\Labeling'$ in which each \candidateText{} is from $\Candidates(\Sites)$.
	Note that we cannot apply the standard method to preserve planarity, i.e., slide labels to the next \candidateText{} and reroute leaders that might cross, as there is no guarantee that our labeling afterwards still respects the given constraints.
	Therefore, it is crucial that we maintain the order of the labels while transforming the labeling.
	This allows us to show that $\Labeling'$ %
	remains planar and, furthermore, respects the constraints because~\Labeling does, i.e., $\Labeling'$ %
	is admissible.

	We iteratively proceed bottom-to-top as follows.
	Let \Label be the bottom-most %
	label %
	that is not yet placed at a canonical \candidateText{}. Furthermore, let $p$ be the port of \Label and $s$ be the site for \Label.
	We move \Label in a non-increasing direction with respect to the length of its leader.
	In particular, if $y(p) < y(s)$, we move~\Label upwards; otherwise, i.e., if $y(p) > y(s)$, we move \Label downwards.
	Note that for $y(p) = y(s)$ the label \Label is already placed at a canonical candidate from $\Candidates(\Sites)$ and we do not need to move it.
	
	We move \Label until it either is positioned at a \candidateText{} from $\Candidates(\Sites)$ or hits another label $\Label'$.
	In the former case, we stop.
	In the latter case, $\Label$ might not yet be at a \candidateText{} from $\Candidates(\Sites)$.
	To ensure that $\Labeling'$ is \admissibleText{}, we ``merge'' these labels into a stack and move from now on this entire stack and thus all its labels simultaneously.
	For a stack, we decide the movement direction based on the majority of the individual movement directions and break ties arbitrarily.
	We move all the not-yet-positioned labels (and stacks) in the same manner.
	If this leads to two stacks touching before they reach \candidatesText{} from $\Candidates(\Sites)$, we merge them and move the single resulting stack as described above.
	Note that since we fix labels bottom-to-top, whenever we hit a stack while moving down, we know that it is already placed at \candidatesText{} from $\Candidates(\Sites)$ and so is thus the merged stack.
	Observe that in the labeling \Labeling, all leaders maintain a vertical distance of at least~$v_{\min}$ to other sites.
	Recall that we place a \candidateText{} $v_{\min}$ away from each site $s$, thus we can never hit a site with our leaders while moving (stacks of) labels, i.e., we do not introduce leader-site crossings. In addition, and since we never swap the order of the labels, we do not introduce leader-leader crossings either.
	Clearly, introducing label-label crossings is also impossible.
	Finally, since we introduce for each site $s$ a \candidateText{} $c$ with $y(c) = y(s)$, this process must eventually stop.
	In the resulting labeling $\Labeling'$, each label is located at a canonical \candidateText{} $c \in \Candidates(\Sites)$ and by above arguments, $\Labeling'$ is planar.
	Observe that the relative order of the labels in~$\Labeling'$ is identical to \Labeling.
	Thus, all constraints are still respected in $\Labeling'$ and the labeling is \admissibleText{}.
\end{proof}
While \cref{lem:sliding-ports-discretization-sufficient} only shows that we preserve admissibility, a closer analysis of the proof and the introduced \candidatesText{} give rise to the following observations.
On the one hand, as already observed by Fink and Suri~\cite{Fink.2016}, we never increase the overall leader length. In particular, for a single label we only decrease the length of its leader and for a stack of labels, we either decrease or maintain the sum of the lengths of the involved leaders.
Thus, for the labeling $\Labeling'$ that we eventually obtain, we know that the sum of the leader lengths can be at most as large as the one in the initial labeling \Labeling, leading to the following corollary.
\begin{corollary}
	\label{cor:sliding-ports-discretization-sufficient-leader-length}
	Let $\Labeling'$ be the labeling obtained from an \admissibleText{} labeling \Labeling, if it exists, as described in \cref{lem:sliding-ports-discretization-sufficient}.
	The leader length of $\Labeling'$ is at most the one of \Labeling.
\end{corollary}
On the other hand, note that for a leader-bend minimal labeling, it only matters whether a site $s$ is labeled at a \candidateText{} $c$ with $y(s) = y(c)$.
Labeling $s$ at any other \candidateText{} $c' \neq c$ contributes one bend to the labeling, independent of the position of $c'$.
While we do not aim for few bends in the procedure described in the proof of \cref{lem:sliding-ports-discretization-sufficient}, we recall that
each site $s \in \Sites$ induces a \candidateText{}{} $c \in \Candidates(\Sites)$ with $y(s) = y(c)$.
Hence, whenever a leader is bendless, every label in the stack it is contained in is placed at a candidate from $\Candidates(\Sites)$.
Thus, if we start in \cref{lem:sliding-ports-discretization-sufficient} with an \admissibleText{} labeling \Labeling with $b$ bends and do not move a stack if it already contains a bendless leader, we can conclude that our procedure does not introduce new bends, or, more formally:
\begin{corollary}
	\label{cor:sliding-ports-discretization-sufficient-leader-bend}
	Let \Instance be an instance of \KSCBLProblemShort{1} with uniform-height labels and sliding \candidatesText{}.
	Let \Labeling be an \admissibleText{} labeling of \Instance, if it exists.
	There exists an \admissibleText{} labeling $\Labeling'$ on the canonical \candidatesText{} $\Candidates(\Sites)$ that has at most the same number of bends as \Labeling does.
\end{corollary}

\cref{lem:sliding-ports-discretization-sufficient} together with \cref{cor:sliding-ports-discretization-sufficient-leader-length,cor:sliding-ports-discretization-sufficient-leader-bend} allows us to show the following theorem.
\begin{theorem}
	\label{thm:sliding-running-time-space} 
	\KSCBLProblemShort{1} for $n$ sites with \emph{uniform-height labels}, $k$ grouping, and~$r$ ordering constraints can be solved in $\BigO{n^{10}\log n + k + \sum_{\Group \in \GroupingConstraints} \Size{\Group}}$ time and \BigO{n^6} space.
\end{theorem}
\begin{proof}
	We transform an instance \Instance of \KSCBLProblemShort{1} for $n$ sites with uniform-height labels and sliding \candidatesText{} into an instance $\Instance'$ of \KSCBLProblemShort{1} with uniform-height labels by introducing the \BigO{n^2} \emph{fixed} canonical \candidatesText{} $\Candidates(\Sites)$ of \Instance.
	\cref{lem:sliding-ports-discretization-sufficient} ensures that we preserve the existence of an \admissibleText{} labeling. 
	Furthermore, due to	\cref{cor:sliding-ports-discretization-sufficient-leader-length} or \cref{cor:sliding-ports-discretization-sufficient-leader-bend}, also the existence of a length- or bend-minimal \admissibleText{} labeling is preserved, respectively.
	
	Since we now have fixed \candidatesText{}, we can use our DP-Algorithm from \cref{sec:fixed-ports}.
	Note that $\Candidates(\Sites)$ can contain \candidatesText{} for which leaders, if placed there, would not satisfy the requirement on a vertical distance of at least $v_{\min}$ to non-incident sites.
	However, recall that we can include this requirement in our DP-Algorithm without increasing its running time.
	Thus, plugging $m = \BigO{n^2}$ into \cref{thm:dp-running-time-space}, we obtain the asymptotic bounds stated in this theorem.
\end{proof}

\section{\KSCBLProblemHeader{Two} is \NP-complete}
\label{sec:two-sided}
\looseness=-1

In the previous section, we showed that \KSCBLProblemShort{1}, while generally %
\NP-hard, can be solved in polynomial time if we have either fixed \candidatesText{} or labels of uniform height.
This does not extend to the generalization of the problem to two sides, as the hardness result presented in this section underlines.
On a high-level, it uses grouping and ordering constraints to divide the vertical space between the two sides of the boundary into independent horizontal strips such that sites can only be labeled inside their respective strip.
By using ordering constraints that run across different strips, we can enforce that sites are labeled on different sides of the boundary.
This %
is the core idea of our reduction, and we describe in the following its construction.

\subparagraph*{Construction of the Instance.}
\looseness=-1
We reduce from \PositiveOneThreeSat, which is a variant of \probname{3-Sat} where each clause contains only positive literals, i.e., the variables appear only non-negated, and a clause is satisfied if and only if exactly one literal per clause evaluates to true.
It is known that this problem is \NP-complete~\cite{Garey.1979,Schaefer.1978}.
Let $\varphi = (\mathcal{X}, \mathcal{R})$ be an instance of \PositiveOneThreeSat, consisting of $N$ variables $\mathcal{X} = \{x_1, \ldots, x_N\}$ and $M$ clauses $\mathcal{R} = \{R_1, \ldots, R_M\}$.
We assume that the left side of the boundary is at $x_l = 0$ and the right side at some $x_r > 0$, where we will fix $x_r$ later but assume for now that it is, without loss of generality, an even integer.
All labels %
have a height of $h = 1$\NewText{,} and we %
leave sufficient vertical space between the \candidatesText{} on the same side of the boundary, such that no two labels can overlap.
\begin{figure}
	\centering
	\includegraphics{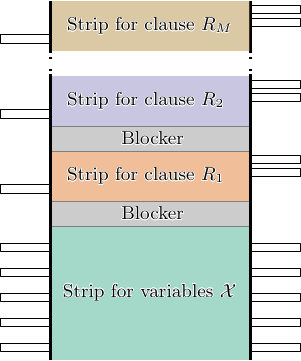}
	\caption{Schematic overview of the constructed instance of \KSCBLProblemShort{2}.} 
	\label{fig:reduction-overview}
\end{figure}
Consider \cref{fig:reduction-overview} for the following description.
From top to bottom, we now assign a horizontal strip that spans the entire span between the two vertical boundaries to each clause, and one for all variables together.
All sites and \candidatesText{} placed for a clause are contained within their strip, while all variable-related sites and \candidatesText{} are contained in the lowest strip.
All strips are pairwise disjoint and separated by blocker gadgets, which we visualize in \cref{fig:illustration-blocker} and describe in the following.
Note that we %
state the concrete placement of the blocker gadgets towards the end of the construction and assume for now that they are placed such that they induce strips of sufficient height.

A \emph{blocker} gadget, %
placed at some $x$-coordinate $x_B$, with $x_l < x_B < x_r$, and $y$-coordinate $y_B$, consists of eight sites.
The three sites shown in \cref{fig:illustration-blocker_a} are placed at $s_1^B = (x_B + M + 1, y_B + 2)$, $s_2^B = (x_B + M + 3, y_B + 3)$, and $s_3^B = (x_B + M + 2, y_B + 4)$.
They form the set $S_L$, and we add the grouping constraint $\{s_1^B, s_2^B, s_3^B\}$ and the ordering constraints indicated in \cref{fig:illustration-blocker_a}, i.e., $s_1^B \Order s_2^B$ and $s_2^B \Order s_3^B$. %
A symmetric set $S_R$ is constructed with the three sites $s_4^B$, $s_5^B$, and $s_6^B$.
We add analogous constraints and place the sites close to the right side of the boundary.
More concretely, we place the sites at $s_4^B = (x_r - x_B - M - 1, y_B + 9)$, $s_5^B = (x_r - x_B - M - 3, y_B + 8)$, and $s_6^B = (x_r - x_B - M - 2, y_B + 7)$.
Furthermore, we add two sites $s_l$ and $s_r$ sufficiently close to the boundary, i.e., at $s_l=(x_B, y_B + 11)$ and $s_r=(x_r - x_B, y_B)$; see \cref{fig:illustration-blocker_b}.
We put $s_l$ in a grouping constraint with the sites of $S_L$ and $s_r$ in a grouping constraint with the sites of $S_R$.
Regarding the \candidatesText{}, we add four on the right side of the boundary, at the $y$-coordinates $y_B + 1$, $y_B + 3$, $y_B + 5$, and $y_B + 11$, and four on the left side of the boundary, at the y-coordinates $y_B$, $y_B + 6$, $y_B + 8$, and $y_B + 10$.
Hence, a blocker gadget consists of eight sites, eight \candidatesText{}, four grouping and four ordering constraints and has a height of eleven without the labels.
Observe that the placement of the sites together with the constraints allows only %
the \admissibleText{} labeling from \cref{fig:illustration-blocker_b} for the blocker. %
As we will select~$x_B$ such that~$s_l$ and $s_r$ are placed close to the respective side of the boundary, a blocker %
partition\NewText{s} the instance by blocking other leaders from passing through it, i.e., it %
creates the aforementioned strips.

\begin{figure}[t]
	\centering
	\begin{subfigure}[t]{.45\linewidth}
		\centering
		\includegraphics{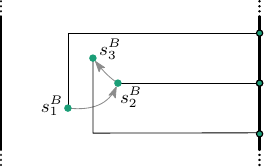}
		\subcaption{Substructure of the blocker gadget.
			We add the grouping constraint $\{s_1^B, s_2^B, s_3^B\}$ and the ordering constraints $s_1^B \Order s_2^B$ and $s_2^B \Order s_3^B$.}
		\label{fig:illustration-blocker_a}
	\end{subfigure}
	\quad
	\begin{subfigure}[t]{.45\linewidth}
		\centering
		\includegraphics{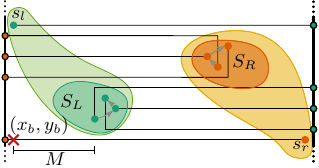}
		\subcaption{The complete blocker gadget and its only \admissibleText{} labeling.
			$S_L$ and $S_R$ are created as in~(\subref{fig:illustration-blocker_a}). We also indicate the point $(x_b, y_b)$ with the red cross and the distance $M$.}
		\label{fig:illustration-blocker_b}
	\end{subfigure}
	\caption{The blocker gadget that subdivides the instance.}
	\label{fig:illustration-blocker}
\end{figure}

\begin{figure}
	\centering
	\begin{subfigure}[t]{.45\linewidth}
		\centering
		\includegraphics[page=1]{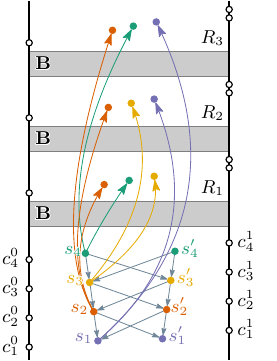}
		\subcaption{The constructed instance.}
		\label{fig:hardness-2-sided-order}
	\end{subfigure}
	\begin{subfigure}[t]{.45\linewidth}
		\centering
		\includegraphics[page=2]{figure_11}
		\subcaption{A (non-\admissibleText{}) labeling that violates an ordering constraint.}
		\label{fig:hardness-2-sided-invalid}
	\end{subfigure}
	\caption{
		The instance created by our reduction for the formula $({x_2} \lor x_3 \lor x_4)\land (x_1 \lor {x_2} \lor x_3)\land ({x_1} \lor {x_2} \lor {x_4})$. Ordering constraints between variables and their occurrences in clauses are indicated in~(\subref{fig:hardness-2-sided-order}).
		If two variables of a clause are set to true, we violate at least one ordering constraint as highlighted with the red leader for the purple site in the clause gadget~$R_3$ in~(\subref{fig:hardness-2-sided-invalid}).
		The labeling from~(\subref{fig:hardness-2-sided-invalid}) induces the variable assignment $x_1 = x_4 = \text{true}$ and $x_2 = x_3 = \text{false}$, which does not satisfy $R_3 = (x_1 \lor x_2 \lor x_4)$ as $x_1$ and $x_4$ are set to true.
	}
	\label{fig:hardness-2-sided}
\end{figure}

We now turn our attention to the bottom-most strip, which hosts the sites and \candidatesText{} for the variables; see also \cref{fig:hardness-2-sided-order} for an illustration of the following construction.
For each variable $x_i$, $1 \leq i \leq N$, we place two \emph{variable-\candidatesText{}} $c_i^0 = (0, 4i - 3)$ and $c_i^1 = (x_r, 4i)$.
Observe that $c_i^0$ is on the left side of the boundary and $c_i^1$ is on the right side of the boundary.
Between them, we place two \emph{variable-sites} $s_i = (x_r / 2 - 2M - i, 4i - 2)$ and $s_i' = (x_r / 2 + 2M + i, 4i - 1)$.
As the sites are vertically between the \candidatesText{}, both sites can use both \candidatesText{}.
The intended meaning is that labeling $s_i$ on the right (left) side of the boundary, the variable $x_i$ should be true (false). 
For each $1 \leq i < N$, we add the ordering constraints $s_{i + 1} \Order s_i$, $s_{i + 1}' \Order s_i$, $s_{i + 1} \Order s_i'$, and $s_{i + 1}' \Order s_i'$ to ensure that the variable-sites are labeled in their vertical order.
As a consequence, the twin site~$s_i'$ must occupy the \candidateText{} $c_i^0$ (or $c_i^1$) that is not used by $s_i$ and thus prevent it from being used by other sites.
Regarding the size of this strip, we observe that we have $2N$ sites and \candidatesText{} for all variables together.
Furthermore, we introduce four ordering constraints per variable-site except for~$s_N$.
This sums up to $4(N - 1)$ ordering constraints.
Regarding the height of the strip, each variable occupies, from $c_i^0$ to $c_i^1$, a height of four on the boundary.
This sums up to $4N$ space on the boundary, excluding the height of the labels.

Finally, for each clause $R_j$, $1 \leq j \leq M$, we place a gadget at a height $y_j$ consisting of three \emph{clause-sites}, $r_j^1$, $r_j^2$, and $r_j^3$, and three \emph{clause-\candidatesText{}}, $c_j^1$,~$c_j^2$, and $c_j^3$.
Each clause-site represents one of the three variables the clause $R_j$ is composed of.
The \candidatesText{} are placed at $c_j^1 = (0, y_j)$, $c_j^2 = (x_r, y_j + 4)$, and $c_j^3 = (x_r, y_j + 5)$, i.e., one \candidateText{} is on the left side and two are on the right side of the boundary.
The sites are placed at $r_j^1 = (x_r / 2 - M + (j - 1), y_j + 1)$, $r_j^2 = (x_r / 2 + (j - 1), y_j + 2)$, and $r_j^3 = (x_r / 2 + M + (j - 1), y_j + 3)$, i.e., vertically between the candidates.
This ensures that for every $1 \leq q \leq 3$, if~$r_j^q$ is labeled on the left side of the boundary, then we can still label both $r_j^{p}$, $1 \leq p \leq 3$, $q \neq p$, on the right side of the boundary.
For every variable $x_i$ in the clause~$R_j$, we add an ordering constraint $s_i \Order r_j^q$ between $s_i$ and the respective clause-site~$r_j^q$ that represents the occurrence of $x_i$ in $R_j$.
This forces the label for $s_i$ to be placed above the label for $r_j^q$ or on the other side of the boundary.
Each clause contributes with three sites and three \candidatesText{} to the size of the instance, which sums up to $3M$ sites and \candidatesText{} for all clauses together.
As each site in a clause gadget has one ordering constraint with the respective variable-site, we also add overall $3M$ ordering constraints.
Regarding the height, we observe that a clause gadget contributes with a height of five to the overall height of the instance.

It is now time to put everything together.
We illustrate the following description in \cref{fig:hardness-2-sided}, which gives an overview of the construction.
Recall that, from top to bottom, each clause gets its own horizontal strip that spans %
between the two vertical boundaries and we have one additional strip for all variables together.
Intuitively, as all strips are separated by blocker gadgets, they prevent any variable-site $s_i$ to be labeled above a clause-site $r_j^q$.
Due to the ordering constraints between variable-sites and their counterparts in the clause gadgets, i.e., due to constraints of the form $s_i \Order r_j^q$, the former sites $s_i$ cannot be labeled on the same side of \Boundary as their latter counterparts $r_j^q$.
As each clause gadget has only one \candidateText{} on the left boundary, only one clause-site per clause can be labeled on the left side of the boundary.
The aforementioned ordering constraint%
$s_i \Order r_j^q$ enforces that %
$r_j^q$ represents a variable $x_i$ whose variable-site $s_i$ is labeled on the right side of the boundary, i.e., %
where $x_i$ is true.
Thus, to not violate an ordering constraint, for each clause %
we must %
have exactly one such variable, %
i.e., one variable that is true and thus satisfies the clause.
To place the blocker gadgets, we recall that the strip for the variables occupies a height of~$4N$ space on the boundary.
We place for every clause~$R_j$, $1 \leq j \leq M$, a blocker at $x_B = j$ and $y_B = 4N + 18(j - 1) + 1$ followed by the clause gadget for~$R_j$ at $y_j = 4N + 18(j - 1) + 13$.
Note that there is sufficient space between the \candidatesText{} in the variable strip, the blockers, and the clause gadgets such that no two labels %
can overlap.
For each clause gadget, we add ordering constraints between the sites from the gadget and those in the blocker immediately below it.
Similarly, for every blocker, we add ordering constraints among the sites from the blocker gadget and the sites in the strip immediately below it.
This forces the sites in the former gadget to be labeled above the sites in the strip below it.
Together, these constraints ensure that sites across different gadgets must be labeled in their vertical order.
Before we can complete the construction, we have to determine the value of $x_r$.
Observe that the variable-sites are placed (horizontally) to the left and right of the clause-sites.
To not interfere, with respect to general position, with the sites for the blocker gadgets, we must select a value for $x_r$ such that $x_r/2 -2M - N$, i.e., the $x$-coordinate of the left-most variable-site, is right of $2M + 3$, i.e., the $x$-coordinate of the right-most site in the set $S_L$ among all blocker-gadgets.
Similarly, $x_r/2 + 2M + N$, i.e., the $x$-coordinate of the right-most variable-site, must be left of $x_r - 2M - 3$, i.e., the $x$-coordinate of the left-most site in the set $S_R$ among all blocker-gadgets.
Setting $x_r = 8M + 2N + 8$ does the trick.
Note that $x_r$ is an even integer, as assumed at the beginning of the construction.
This completes the construction and we are now ready to show \cref{thm:hardness-2-sided}.

\begin{restatable}{theorem}{hardnessTwoSidedTheorem}
	\label{thm:hardness-2-sided}
	Deciding if an instance of \KSCBLProblemShort{2} has an \admissibleText{} labeling is \NP-complete, even for uniform-height labels and fixed \candidatesText{}.
\end{restatable}
\begin{proof}	
	\looseness=-1
	We argue \NP-containment and \NP-hardness separately.
	
	\proofsubparagraph*{\NP-containment.}
	Containment in \NP{} follows immediately since for any assignment of the $n$ sites to the $m$ candidates we can check in polynomial time whether the leaders are crossing free and no grouping and ordering constraints are violated.
	Note that we have fixed \candidatesText{}.
	Hence, the precision for the coordinates of the sites, \candidatesText{}, and thus also leaders, that we need when checking that a given labeling is admissible is determined by the (finite) precision of these values in the input. 
	
	\proofsubparagraph*{\NP-hardness.}
	Let $\varphi = (\mathcal{X}, \mathcal{R})$ be an instance of \PositiveOneThreeSat, consisting of $N$ variables $\mathcal{X} = \{x_1, \ldots, x_N\}$ and $M$ clauses $\mathcal{R} = \{R_1, \ldots, R_M\}$.
	Furthermore, let $\Instance(\varphi)$ be the instance of \KSCBLProblemShort{2} obtained with the construction from above.
	We observe that we create one blocker for each of the $M$ clauses, i.e., we have in total $M$ blocker gadgets in the construction.
	Hence, $\Instance(\varphi)$ consists of $11M + 2N$ sites and \candidatesText{}.
	Regarding the constraints, we have~$4M$ grouping and $4M$ ordering constraints from the blockers.
	Furthermore, we introduced $4(N - 1)$ ordering constraints among the variables, $3M$ ordering constraints among clause-sites and variable-sites, and $11M$ ordering constraints to maintain the vertical order among sites from different gadgets.
	To bound the overall height of the instance, we observe that we place the last clause gadget at the $y$-coordinate $4N + 18(M - 1) + 13$.
	As it has a height of five, we conclude that the height of the instance, without the labels, is $4N + 18M$.
	The width of the instance is equal to $x_r$, which we set to $8M + 2N + 8$.
	Finally, since we only used integer values for the coordinates of the sites and \candidatesText{}, we conclude that we can construct $\Instance(\varphi)$ on an integer grid of a size that is bounded  polynomially in the size of $\varphi$.
	Furthermore, we can readily observe that $\Instance(\varphi)$ is in general position and can be created in polynomial time with respect to the size of $\varphi$.
	Recall \cref{fig:hardness-2-sided} for the following correctness arguments.
	
	\proofsubparagraph*{($\boldsymbol{\Rightarrow}$)}
	Assume that $\varphi$ is a positive instance of \PositiveOneThreeSat.
	Hence, there exists a truth assignment $\Phi\colon \mathcal{X} \to \{0, 1\}$ such that for each clause $R_i \in \mathcal{R}$, $1 \leq i \leq M$, we can find a variable $x_j \in R_i$ so that $\Phi(x_j) = 1$ and $\Phi(x) = 0$, for all $x \in R_i \setminus \{x_j\}$, i.e., exactly one literal of each clause evaluates to true under $\Phi$.
	We replicate this assignment in a labeling \Labeling of $\Instance(\varphi)$ as follows.
	For $1 \leq i \leq N$, if $\Phi(x_i) = 1$ we label $s_i$ at the candidate on the right side of the boundary, i.e., set $\Labeling(s_i) = c_i^1$, and $s_i'$ consequently on the left side of the boundary, i.e., set $\Labeling(s_i') = c_i^0$.
	If $\Phi(x_i) = 0$, we mirror the assignment, i.e., set $\Labeling(s_i) = c_i^0$ and $\Labeling(s_i') = c_i^1$.
	We label a site in a clause gadget on the left boundary if the corresponding variable it represents satisfies the clause, i.e., is true, and otherwise on the right side of the boundary.
	For the sites that make up the blocker gadgets, we label them according to \cref{fig:illustration-blocker_b}.
	What is left to do is argue that~\Labeling is \admissibleText{}.
	The labeling of the blocker gadgets is planar by construction.
	As for any variable $x \in \mathcal{X}$, we have either $\Phi(x) = 1$ or $\Phi(x) = 0$, but never both, the label positions for the sites representing the variables are well-defined.
	Furthermore, as $c_i^1$ is above $s_i$ and $s_i'$, and $c_i^0$ below~$s_i$ and $s_i'$, any possible variable assignment $\Phi$ can be transformed into a planar labeling.
	Finally, since we placed the \candidatesText{} and sites with sufficient space from each other, mimicking $\Phi$ in \Labeling as described above %
	results in a planar labeling.
	For the sites in the clause gadgets, as $\Phi$ ensures that exactly one literal evaluates to true, we know that one site %
	is labeled on the left boundary and the other two on the right boundary.
	This is exactly the distribution of the three \candidatesText{} we have chosen when creating the clause gadgets; see also \cref{fig:hardness-2-sided}.
	Furthermore, we have ensured that no matter which site we label on the left side, there is a way to label the remaining two sites on the right side.
	Therefore,~\Labeling is planar.
	To show that \Labeling respects also the constraints, we first note that we respect the constraints in the blocker gadgets by construction.
	Furthermore, for any $1 \leq i < N$ we have $y(c_i^0) \leq y(c_i^1) \leq y(c_{i + 1}^0) \leq y(c_{i + 1}^1)$.
	Hence, the ordering constraints among the vertices for the variables and their twins are satisfied.
	Similarly, since we label each clause gadget within its slice, and all sites in a blocker at the \candidatesText{} created for it, we label the sites in their vertical order and thus also satisfy the ordering constraints among sites of different gadgets.
	Hence, we only have to consider the ordering constraints between variable-sites and their respective occurrence in clauses.
	Let the variable $x_i$ appear in the clause $R_j$ and let $s_i \Order r_j^q$, $q \in \{1,2,3\}$, be the ordering constraint that puts them into relation.
	Since $s_i$ represents a variable and $r_j^q$ its occurrence in a clause, we know that $s_i$ %
	is labeled below $r_j^q$ in \Labeling due to the blockers.
	Therefore, to respect $s_i \Order r_j^q$, we must show that $s_i$ and $r_j^q$ are labeled on different sides of the boundary.
	There are two cases: $\Phi(x_i) = 0$ or $\Phi(x_i) = 1$.
	In the former case, we label $s_i$ on the left side and, as the clause~$R_j$ is then not satisfied by the variable~$x_i$,~$r_j^q$ is labeled on the right side.
	In the latter case, i.e., if $\Phi(x_i) = 1$, we label~$s_i$ on the right side and, as it satisfies the clause, $r_j^q$ is labeled on the left side.
	As we label them in both cases on different sides of the boundary, this ordering constraint is trivially satisfied.
	Since we selected $s_i \Order r_j^q$ arbitrarily, we know that this holds for all constraints among sites for variables and clauses and thus is~\Labeling \admissibleText{}.
	
	\proofsubparagraph*{($\boldsymbol{\Leftarrow}$)}	
	\looseness=-1
	Let $\Instance(\varphi)$ possess an \admissibleText{} labeling \Labeling.
	Recall that we have in $\Instance(\varphi)$ as many sites as \candidatesText{}.
	\Labeling respects the ordering constraints among the sites introduced for the variables and for sites from different gadgets, which implies, due to the transitive nature of ordering constraints and the existence of the twin $s_i'$ for each $s_i$, that we label all sites in their relative vertical order and in particular $s_i$ below $s_{i + 1}$ for $1 \leq i < N$ and $s_N$ below all blocker and clause gadgets.
	Furthermore, the blocker gadget right above the sites for the variables prevents any of those sites to be labeled at a \candidateText{} above $c_N^1$.
	Combining all these observations, we conclude that $s_i$, $1 \leq i \leq N$, must be labeled at $c_i^0$ or $c_i^1$ in \Labeling, i.e., we have $\Labeling(s_i) \in \{c_i^0, c_i^1\}$.
	We now create a truth assignment $\Phi\colon \mathcal{X} \to \{0, 1\}$ over $\mathcal{X}$ based on \Labeling.
	More concretely, if we have $\Labeling(s_i) = c_i^1$ we set $\Phi(x_i) = 1$.
	Similarly, if we have $\Labeling(s_i) = c_i^0$ we set $\Phi(x_i) = 0$.   
	As argued above, due to the structure of $\Instance(\varphi)$ we have in \Labeling each $s_i$ labeled at one of those two \candidatesText{}.
	Thus, $\Phi$ is well-defined.
	All that remains to do is to show that in every clause $R_j$, $1 \leq j \leq M$, exactly one literal evaluates to true.
	We remind the reader that every literal in $\varphi$ is a non-negated variable.
	The labeling~\Labeling respects all the constraints of $\Instance(\varphi)$, in particular the ordering constraints among sites of clauses and variables.
	However, due to the blocker gadgets, especially the one between the variable slice and the first clause gadget, in any \admissibleText{} labeling, and therefore also in \Labeling, all sites in the variable slice are labeled below those in the clause gadgets.
	Therefore, to respect the above-mentioned constraints, \Labeling must satisfy them ``trivially'', i.e., by labeling the sites on different sides of the boundary.
	Hence, for each site $s_i$ in the variable slice that we label on the right side, i.e., at $c_i^1$, we must label all sites for its occurrences in clauses on the left side, as there is a corresponding constraint that enforces this.
	A symmetric argument holds if we label $s_i$ on the left side.
	However, since for every clause gadget representing some clause $R_j$, $1 \leq j \leq M$, we create three ordering constraints and there is only one \candidateText{} on the left side and two \candidatesText{} on the right side, we know that \Labeling can only be \admissibleText{} if exactly one of the three sites that make up the clause gadget is labeled on the left side, i.e., the variable it represents is considered true.
	Thus, $\Phi$ satisfies exactly one literal of each clause in $\varphi$, i.e., $\varphi$ is a positive instance of \PositiveOneThreeSat.
\end{proof}

\section{Experimental Evaluation}
\label{sec:experiments}
\looseness=-1
From a theoretical point of view, \cref{thm:dp-running-time-space,thm:sliding-running-time-space} show that \KSCBLProblemShort{1} can be solved in polynomial time if we assume fixed \candidatesText{} or uniform-height labels.
However, the large (asymptotic) running time of the algorithms calls into question their practical relevance.
Given that the problem is motivated from real-world applications, we now seek to experimentally evaluate the approaches on real-world instances.
We evaluate implementations of our algorithms for both the fixed and sliding \candidatesText{} setting.
Our algorithm for the sliding \candidatesText{} setting necessitates even for a small instance of just 10 features a large number of over 600 \candidatesText{}.
As a result, the algorithm did not terminate within a 2 hour time window, seemingly confirming the practical infeasibility suggested by the theoretical runtime.
We, therefore, focus in the following on evaluating our algorithm for the fixed \candidatesText{} setting, which does not require these large numbers of \candidatesText{} per feature and therefore yields better runtime results.

\begin{figure}
	\centering
	\includegraphics[page=1]{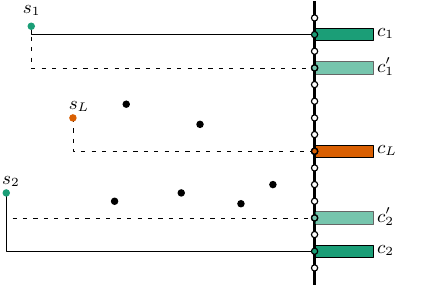}
	\caption{No matter if we have the sub-instance $I = (s_1, c_1, s_2, c_2)$ or $I' = (s_1, c_1', s_2, c_2')$, if we label $s_L$ at $c_L$, the leader $\Leader_L = (s_L, c_L)$ %
		either respects the constraints or does not respect them in both sub-instances.} 
	\label{fig:avoid-respect-constraints}
\end{figure}

\subparagraph*{Implementation Details.}
\looseness=-1
Our algorithm is implemented in C++17 and computes a leader-length minimal labeling on fixed \candidatesText{}.
Although it could handle non-uniform label heights, we settled for labels with a uniform height of $20$ pixels.
Note that this only affects the existence of an \admissibleText{} labeling, but has no influence on the theoretic running time of the algorithm that we stated in \cref{thm:dp-running-time-space}.
We build the implementation of our PQ-A-graph on a \emph{PC-Tree}\footnote{PC-Trees can be seen as a variant of PQ-trees, and we refer to Hsu and McConnell~\cite{Hsu.2003} for a formal definition of PC-Trees. Furthermore, it is known that we can simulate PQ-trees using PC-Trees~\cite{Haeupler2008}.} implementation by Fink, Pfretzschner, and Rutter~\cite{Fink.2021} that outperforms existing PQ-tree implementations.
In addition, we use a range tree implementation by Weihs~\cite{Weihs.2020} that was already used in the literature~\cite{Weihs.2018}.
We pre-compute the least common ancestor of all pairs of leaves to not re-compute it several times.
To avoid checking multiple times for similar sub-instances whether the leader for the leftmost site respects the constraints, we use the following observation.
Consider the %
setting from \cref{fig:avoid-respect-constraints}, where the bounding box of the sites in a sub-instance $I = (s_1, c_1, s_2, c_2)$ is identical to the one in a sub-instance $I' = (s_1, c_1', s_2, c_2')$.
As we assume general position, this bounding box describes the same set of sites and both instances are bounded by leaders originating from $s_1$ and $s_2$.
Hence, everything that affects the outcome of \RespectsConstraints{I, \PQAGraph, \Leader_L = (s_L, c_L)} is identical, given that we have the \candidateText{}~$c_L$ in $I$ and $I'$. %
This idea is inspired by the practical considerations made by Niedermann et al.~\cite{Niedermann.2017} to speed up their contour labeling algorithm.
Above observations give rise to the use of memoization to avoid re-running \RespectsConstraints{I, \PQAGraph, \Leader_L} and instead (re-)use the result from \RespectsConstraints{I', \PQAGraph, \Leader_L}.

In addition to our DP, we use an existing ILP-formulation for boundary labeling on fixed \candidatesText{}~\cite{Barth.2019} as a reference for the running time.
Note that this should only simulate a ``textbook'' implementation of a labeling algorithm and does not claim to be the most efficient way to obtain a leader-length minimal labeling on fixed \candidatesText{}.
Finally, this \emph{Na\"ive-ILP} is unaware of our constraints.
\begin{table}
	\centering
	\caption{Properties of the instances from the dataset human anatomy that is based on illustrations from the Sobotta atlas of human anatomy~\cite{Waschke.2013}.}
	\begin{threeparttable}
		\begin{tabularx}{\linewidth}{lrrrrX}
			\toprule
			\multicolumn{1}{c}{Figure} & \multicolumn{1}{c}{$n$} & \multicolumn{1}{c}{$m$} & \multicolumn{1}{c}{$k$} & \multicolumn{1}{c}{$r$} & Definition of the constraints \\
			\midrule
			Fig.~8.30 & 11 & 44 & 3 & 0 & Groups based on colored regions or curly brackets enclosing labels.\\
			Fig.~8.81 &  9 & 60 & 0 & 7 & A nerve branching off another nerve should be labeled below its ``parent''.\\
			Fig.~9.23 & 11 & 53 & 3 & 0 & Overlapping groups based on explicit curly brackets or colored regions.\\
			Fig.~12.33 &  9 & 38 & 3 & 0 & Grouping based on colored regions. Sites on the boundary of two regions are in both regions, i.e., groups overlap.\\
			Fig.~12.59 & 19\tnote{*} & 62 & 3 & 6 & Grouping based on curly brackets, ordering based on Roman letters next to some labels.\\
			\bottomrule
		\end{tabularx}
		\begin{tablenotes}
			\item[*] The original figure contains 31 sites labeled on the left and right side of the illustration. However, we took only the sites labeled on the left side.
		\end{tablenotes}
	\end{threeparttable}
	\label{tab:experiments-book-properties}
\end{table}

\begin{figure}
	\centering
	\begin{subfigure}[t]{.45\linewidth}
		\centering
		\includegraphics[width=\linewidth]{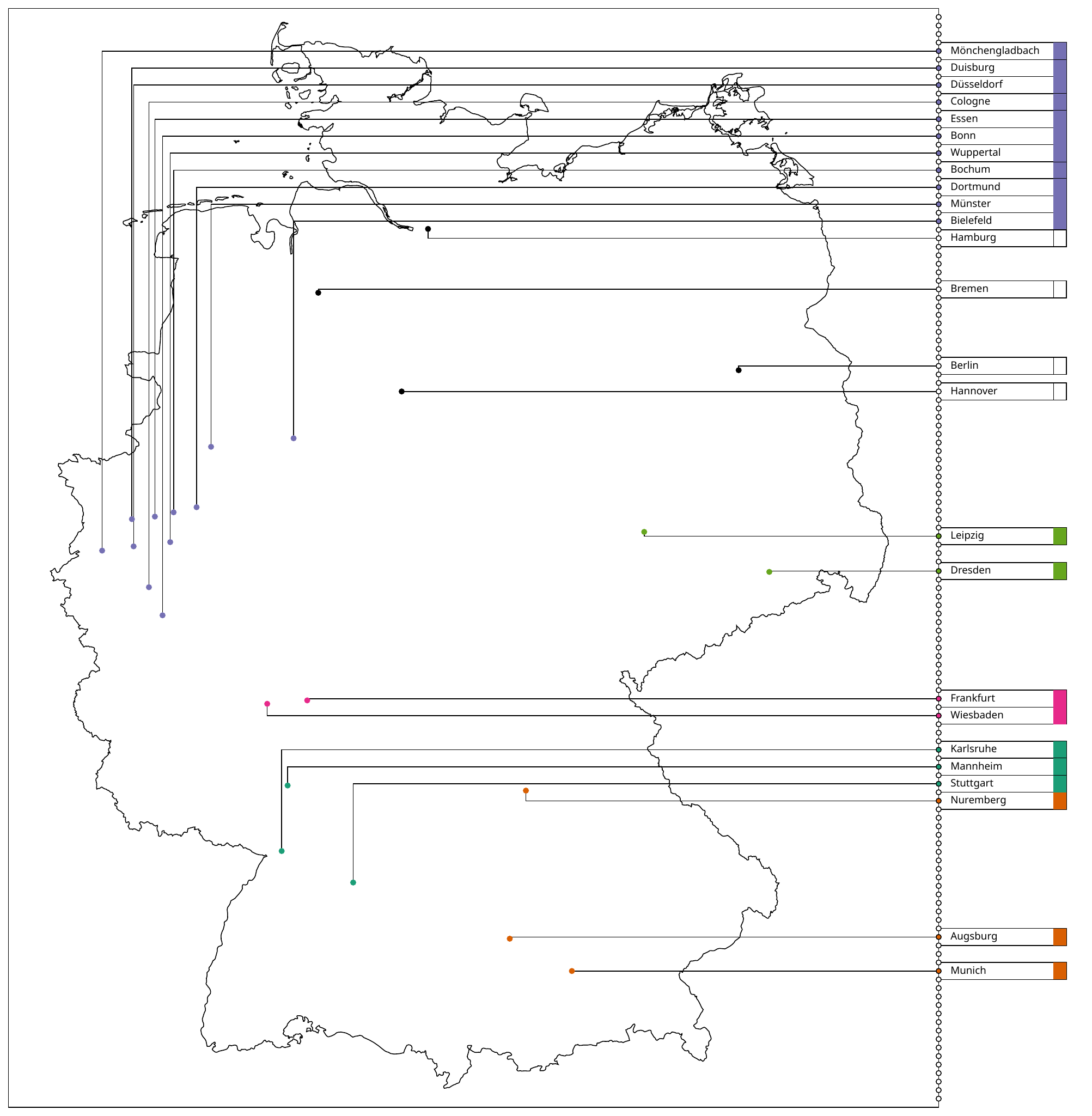}
		\subcaption{The 25 largest cities of Germany.}
		\label{fig:example-dataset-cities}
	\end{subfigure}
	\begin{subfigure}[t]{.45\linewidth}
		\centering
		\includegraphics[width=\linewidth]{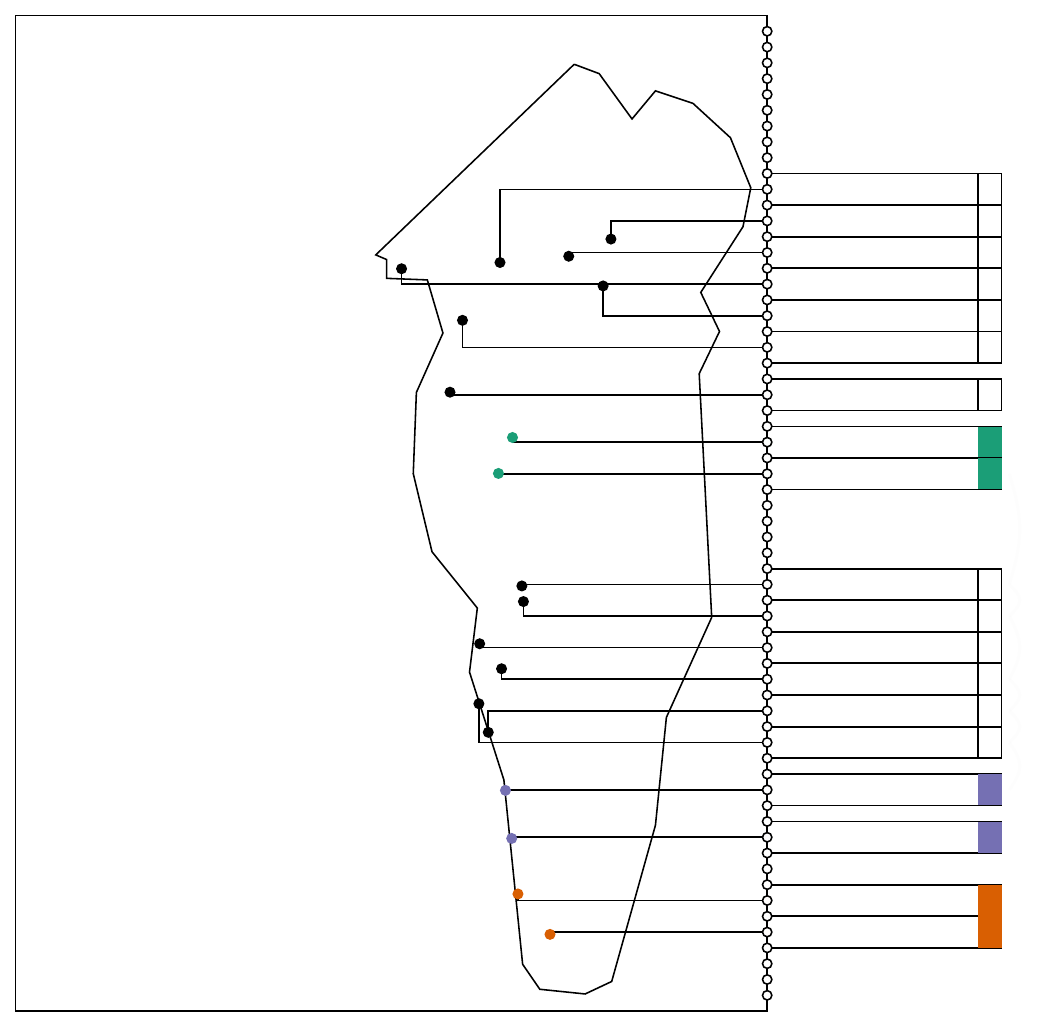}
		\subcaption{Figure~12.59 adapted from the Sobotta atlas of human anatomy~\cite{Waschke.2013}.}
		\label{fig:example-dataset-human-anatomy}
	\end{subfigure}
	\caption{
		Sample instances of our datasets together with labelings computed by our algorithm.
		\CandidatesText{} are placed every ten pixels in both instances.
		Colors indicate sites of the same group and arrows next to the labels visualize ordering constraints. Figures not to scale.
	}
	\label{fig:example-dataset}
\end{figure}

\subparagraph*{Datasets.}
We crafted two datasets based on real-world data.
They are available on OSF~\cite{OSF.CBL.2025}.
The first dataset, \emph{cities}, contains instances with the $n \in \{10, 15, 20, 25, 30, 35, 40, 45\}$ largest cities from Austria, Germany, and Italy, respectively, obtained from \texttt{simplemaps.com}~\cite{SimpleMaps.com.2023}.
For each combination of country and $n$, we create four instances: one without constraints, which we use for the ILP, and three where we group the cities according to the administrative regions of the respective country.
In addition, in one of the instances that contains the grouping constraints, we order the cities in each group according to their administrative status (create so-called \emph{intra-group} constraints).
The fourth instance is created by ordering the groups according to their population computed from the cities in the instance (create so-called \emph{inter-group} constraints).
To do this, we select a representative site from each group and insert the corresponding ordering constraints.
This results in at most 13 grouping constraints.
We create three variants of this dataset that differ in the number of \candidatesText{}.
In \emph{cities-2x}, each instance has $m = 2n$ many \candidatesText{}, and \emph{cities-90} defines 90 \candidatesText{} per instance.
For these two variants, we maintain a distance of at least twenty pixels, the label height $h$, between two \candidatesText{}.
For \emph{cities-10px}, we perform differently.
We take an initial height for the boundary and place a \candidateText{} every ten pixels.
If this leads to too few \candidatesText{}, we increase the height of the boundary and add more \candidatesText{} accordingly.
This approach led to 71, 128, and 130 \candidatesText{} for the Austrian, German, and Italian cities, respectively, independent of the number of cities in the instance.
We show in \cref{fig:example-dataset-cities} a sample instance of cities-10px together with a labeling.

The second dataset, \emph{human anatomy}, contains five instances obtained from the Sobotta atlas of human anatomy~\cite{Waschke.2013} that Niedermann et al.~used to evaluate the performance of their labeling algorithm~\cite{Niedermann.2017}.
These instances are enriched with sometimes overlapping grouping and ordering constraints as described in \cref{tab:experiments-book-properties}.
Note that the book uses contour labeling for their illustrations.
We have therefore selected the instances such that the labeling from the book suggests the existence of an \admissibleText{} labeling within our model.
In particular, for these instances the labeling from the book either has the labels on one side of the contour only, or there is enough space on one side so that labels from the other side can be placed there.
Similar to Niedermann et al., we place a \candidateText{} every ten pixels.
\cref{fig:example-dataset-human-anatomy} shows an instance of this dataset together with a labeling.

\subparagraph*{Experimental Setup.}
All %
instances from the cities dataset were solved on a compute cluster with Intel Xeon E5-2640 v4 10-core processors at 2.40GHz that have access to 160 GB of RAM.
We set a hard memory limit of 96 GB that was never exceeded.
To simulate a real-world setting, we computed the instances from the human anatomy dataset on an off-the shelf laptop with an Intel Core i5-8265U 4-core processor at 1.60GHz.
There, we had, inside a WSL2 environment, access to 7.8 GB of RAM.

\subparagraph*{Results.}
The experiments revealed that for the cities dataset only every second instance was feasible, i.e., admitted an \admissibleText{} labeling.
In particular, those instances that contain grouping and ordering constraints were often infeasible.
Comparing the feasible instances with both types of constraints to their counterpart without ordering constraints revealed that the former have a total leader length that is on average less than one percent larger.
This negligible increase gives the impression that ordering constraints are well suited for enforcing locally limited orders but should not be used to put labels for sites far away into relation.
All instances from the human anatomy dataset were feasible.
Based on these results, we conclude regarding the feasibility of the instances that we should not only consider the semantics of sites but also their (geometric) position when defining constraints.
\cref{fig:example-dataset} shows two sample labelings from our experiments and we refer to our OSF repository~\cite{OSF.CBL.2025} for the labelings of all feasible instances.

\begin{figure}
	\centering
	\begin{subfigure}[t]{0.98\linewidth}
		\centering
		\includegraphics[width=\linewidth]{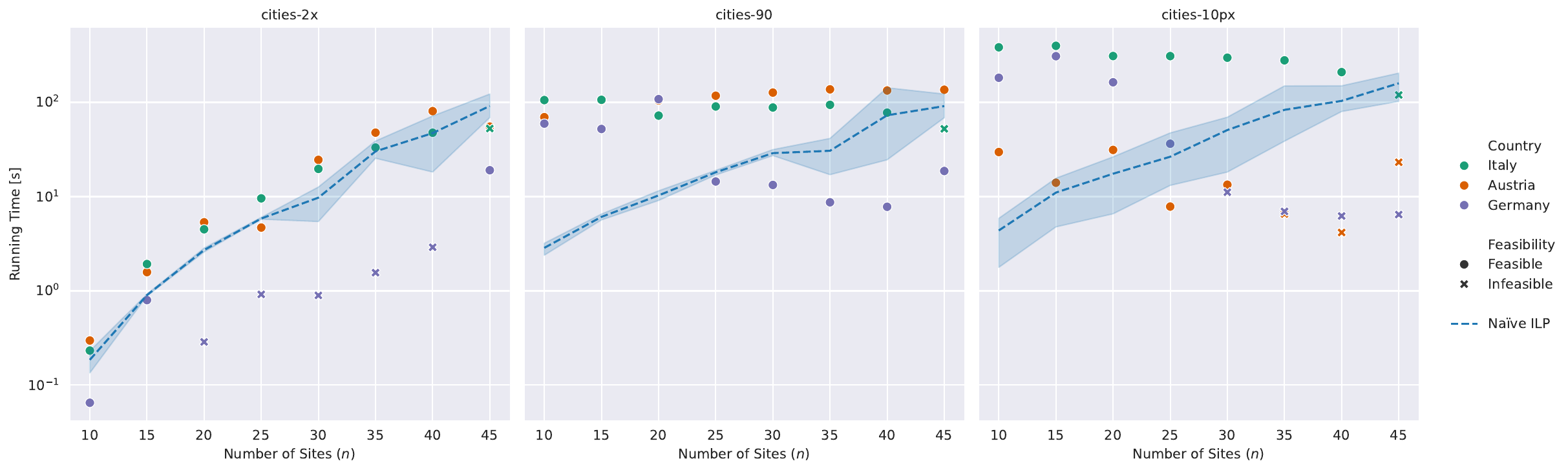}
		\subcaption{Running time for instances that only contain grouping constraints.}
		\label{fig:running-time-cities-grouping}
	\end{subfigure}
	
	\begin{subfigure}[t]{0.98\linewidth}
		\centering
		\includegraphics[width=\linewidth]{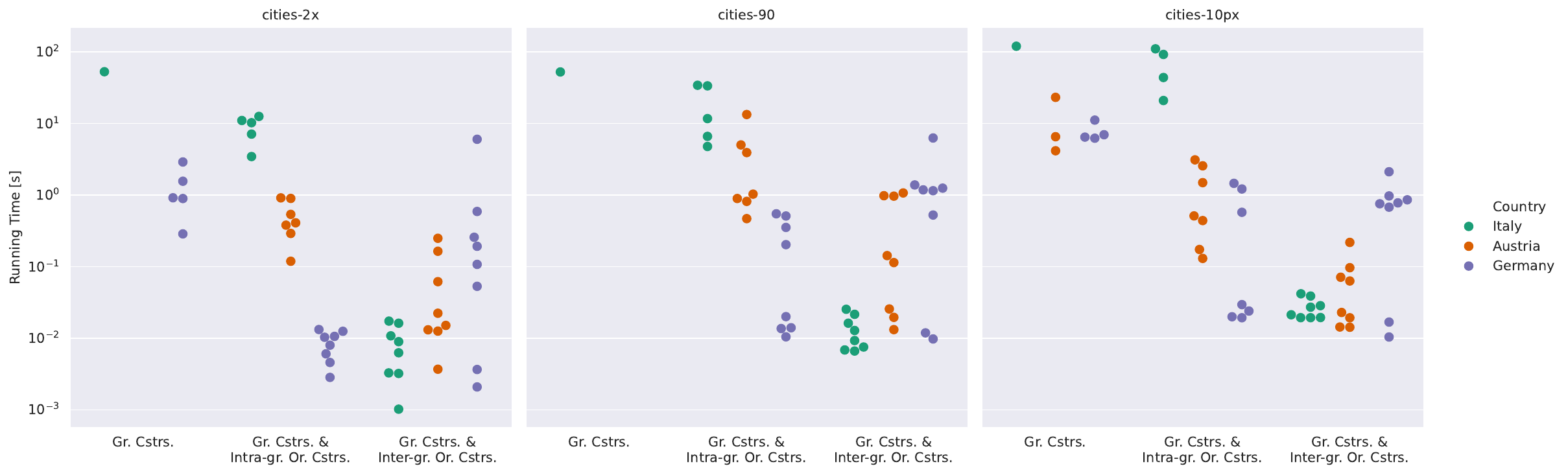}
		\subcaption{Time required to detect infeasible instances.}
		\label{fig:running-time-cities-infeasible}
	\end{subfigure}
	\caption{Running time (in seconds) on the cities datasets ($\log$-plots). \emph{Gr.} stands for grouping and \emph{Or.} for ordering constraints (abbreviated as \emph{Cstrs.})}
	\label{fig:running-time-cities}
\end{figure}

We present in \cref{fig:running-time-cities,tab:running-time-book} the measured running times.\footnote{We measured the wall-clock time, excluding the reading (and parsing) of the instance from and the writing of the labeling to the disk, but including any other preprocessing steps.}
For the cities dataset, we report in \cref{fig:running-time-cities} the running time plots.
Recall that the number of \candidatesText{} in cities-2x depends on the number of sites and thus varies between different data points.
For the other variants of this dataset, the number of \candidatesText{} is the same across all instances from the same country.
As many of the instances with grouping \emph{and} ordering constraints were infeasible, we refrain from plotting their running times in detail and show in \cref{fig:running-time-cities-grouping} only the running times for the instances with grouping constraints.
We also plot in \cref{fig:running-time-cities-grouping} the average running time of the ILP together with the range of measured running times as reference.
\cref{fig:running-time-cities-infeasible} shows the running times of all infeasible instances.
\cref{tab:running-time-book} contains the running times for our solver and the ILP on the instances taken from the Sobotta atlas of human anatomy.
\begin{table}
	\centering
	\caption{Running time (in seconds) of our algorithms on the human anatomy instances.}
	\label{tab:running-time-book}
	\begin{threeparttable}
		\begin{tabular}{lrrr}
			\toprule
			&&\multicolumn{2}{c}{Our DP-Algorithm}\\
			\cmidrule{3-4}
			\multicolumn{1}{c}{Instance} & \multicolumn{1}{c}{\SolverILP} & \multicolumn{1}{c}{Total} & \multicolumn{1}{c}{of which bookkeeping\tnote{*}} \\
			\midrule
			Fig.~8.30  & 0.7922 &   0.6084 & 0.0003\\
			Fig.~8.81  & 0.4295 &   0.3700 & 0.0003\\
			Fig.~9.23  & 0.5023 &   0.3658 & 0.0004\\
			Fig.~12.33 & 0.2519 &   0.0418 & 0.0003\\
			Fig.~12.59 & 2.4632 &   3.6388 & 0.0006\\
			\bottomrule	    
		\end{tabular}
		\begin{tablenotes}
			\item[*] This includes everything but filling the DP-Table $D$.
		\end{tablenotes}
	\end{threeparttable}
\end{table}

From the measurements, we can observe that our algorithm is for small to medium-sized instances fast enough to compute a labeling in a few seconds or classify instances as infeasible, even on an off-the-shelf laptop (\cref{tab:running-time-book}).
On the other hand, if the instance is large, the running time can be seen as a limitation of our approach.
In particular, we could measure a running time of up to seven minutes for the largest instances (\cref{fig:running-time-cities-grouping}).
On a positive note, infeasibility was detected within ten seconds, even for many large instances (\cref{fig:running-time-cities-infeasible}).
We observed that the running time seems to be more dependent on the number of \candidatesText{} than on the number of sites.
This observation is consistent with the findings we made for the setting with sliding \candidatesText.
Finally, we noticed that the feasibility of the instances and the quality of the computed labelings strongly depend on the position of the \candidatesText{}.
Therefore, we think that the study of sliding \candidatesText{} is also interesting from an applied perspective, despite the high running times that we could measure.

\section{Conclusion}
We introduced and studied grouping and ordering constraints in boundary labeling,
which occur in real-world labelings; recall \cref{fig:illustration-motivation}.
While finding an \admissibleText{} labeling is %
\NP-hard, polynomial-time algorithms for one-sided instances with fixed \candidatesText{} or uniform-height labels exist.
Our experiments have shown that our algorithm for fixed \candidatesText{} has a reasonable running time for small instances.
Still, future work could try to speed up the admissibility checks in our dynamic program to reduce its overall running time.

Grouping constraints enforce labels to be on the same side of the boundary, while ordering constraints, at the same time, only apply between labels that actually are on the same side.
While this interpretation is general enough to apply to many settings and finds justification in real-world labelings, it is of course not the only way to model our constraints.
In particular, one could study grouping constraints that apply to each of the two sides independently, or ordering constraints that also constrain the relative placement of labels across different sides of the boundary.
We find the latter variant in particular interesting, since our reduction heavily relies on ordering constraints being only relevant among labels on the same side.
Furthermore, both hardness constructions use overlapping grouping constraints.
While they have their merit, they are less common in real-world labelings.
Hence, we see obtaining polynomial-time algorithms for disjoint grouping constraints a relevant quest for future work.

Incorporating semantic \emph{soft} constraints, i.e., considering the task of maximizing the number of satisfied constraints in a planar labeling, is also an interesting direction for further investigation.
Since we can also label features other than points, 
it is  worth studying a variant of this problem with uncertain or variable site locations.
Similarly, the support of other leader types or entire other external labeling styles should be investigated.
Finally, the visual quality of the produced labeling should be experimentally and empirically evaluated, e.g., with a user study.

\bibliography{references}

\end{document}